\documentclass{aastex}
\usepackage{spr-astr-addons}
\usepackage{natbib}
\usepackage{url}

\RequirePackage{color}

\begin{document}

\title{Variable stars in the field of the Galactic bulge globular cluster NGC 6522}
\slugcomment{Not to appear in Nonlearned J., 45.}
\shorttitle{Variable stars in the field of NGC 6522}
\shortauthors{Arellano Ferro et al.}

\author{A. Arellano Ferro\altaffilmark{1}}  
\author{Z. Prudil\altaffilmark{2}}
\author{M.A. Yepez\altaffilmark{1}}
\author{I. Bustos Fierro\altaffilmark{3}}
\author{A. Luna\altaffilmark{2,4}}

\altaffiltext{1}{Instituto de Astronom\'ia, Universidad Nacional Aut\'onoma de M\'exico. Ciudad Universitaria CP 04510, Mexico: (armando@astro.unam.mx).}
\altaffiltext{2}{European Southern Observatory, Karl-Schwarzschild-Straße 2, 85748, Garching, Germany: (Zdenek.Prudil@eso.org), (alonso.luna@eso.org).}
\altaffiltext{3}{Observatorio Astronómico, Universidad Nacional de Córdoba, Córdoba C.P. 5000, Argentina:(ivan.bustos.fierro@unc.edu.ar).}
\altaffiltext{4}{Instituto de Astrofísica, Facultad de Ciencias Exactas, Universidad Andrés Bello, Fernández Concha 700, Las Condes, Santiago, Chile.}


\begin{abstract}

We present a comprehensive analysis of the variable stars projected on the field of the Galactic bulge globular cluster NGC 6522, offering valuable insights into their characteristics. Using proper motions from Gaia DR3, we aim to distinguish between field stars and true cluster members. For an accurate color-magnitude diagram of the member stars, we produced a differential reddening map. We detect and discuss the peculiarities of variable stars of the types RR Lyrae, type II Cepheids, Long Period Variables (LPV) and eclipsing binaries, whose light curves are available through the OGLE III and IV databases. Notably, we explore the variable V24, which shows a prominent phase modulation resulting from period changes in a time scale of ~1100 days. The variable stars among the cluster members serve as indicators of the cluster metallicity and distance; these determinations are based on their light curves. With the Fourier light curve decomposition of three RRc stars, we have derived the following cluster parameters: the metallicity in the spectroscopic scale [Fe/H]$_{\rm UVES}$=--1.16 $\pm$0.09; and the mean distance $D=8.77  \pm 0.16$ kpc.
\end{abstract}


\section{Introduction}
The globular cluster NGC~6522 (C1800-300 in the IAU nomenclature) ($\alpha = 18^{\mbox{\scriptsize h}}
03^{\mbox{\scriptsize m}} 34.08^{\mbox{\scriptsize s}}$, $\delta = -30\degr 02\arcmin
02.7\arcsec$, J2000; $l = 1.02\degr$, $b = -3.93\degr$) is one among some 67 
globular clusters projected against the Galactic bulge and 85 candidate clusters within 10 degrees around the Galactic center \citep{Bica2019}. It is a core collapsed cluster with 0.05 arcminute core radius,
1.0 arcminute half-light radius \citep{Harris1996} (2010 edition) and it has been argued  to be among the oldest clusters in the Galaxy, about 2 Gyrs older than old M5 and 47 Tuc \citep{Barbuy2009}.
It is affected by very large 
reddening, $E(B-V)\sim 0.48-0.58$ mag, as it has been estimated by several authors and dust map calibrations \citep{Harris1996,Schlegel1998,Schlafly2011}.
Being a bulge cluster, its field is highly contaminated by field stars, and so it is the variable star population projected against the cluster field. 
Many variable stars have been detected near NGC~6522 by the $Gaia$ mission \citep{Gaia2016} and the Optical Gravitational Lensing Experiment (OGLE) \citep{Udalski1992}.
 While the employment of RR Lyrae stars as distance and metallicity indicators in globular clusters via the Fourier decomposition of their light curves has been amply demonstrated \citep{Arellano2023}, using  variables as distance and metallicity indicators, requires a detailed  membership analysis and a critical evaluation of the positions in the Colour Magnitude diagram (CMD) of all the detected variables in the field of the cluster. Once we identify the most likely cluster member variables, we are in position to discuss their implied physical parameters. In the present paper we perform this analysis and report our conclusions on what seems the authentic population of variables in NGC 6522. We shall compare our best estimates of the mean cluster [Fe/H] and distance with the spectroscopic values obtained by \citet{Barbuy2009} and the distances compilation from
the literature performed by \citet{Baumgardt2021}. In the process, we refined the variables ephemerides, their classification, highlight the detected Blazhko variables and determined their modulation periods. We also include a discussion of a deferentially dereddened CMD  based on the differential reddening map of \citet{AlonsoGaria2012}. 

\section{Employed photometric data}
\label{sec:DATA}

Photometric \emph{VI} data for a large number of stars in the grand field around NGC 6522 are available in the OGLE III \citep{Soszynski2013} and OGLE IV \citep{Soszynski2014}. We have sustained nearly all our photometric analysis on these data. On the other hand positional and proper motions data are available in the $Gaia$-DR3 \citep{Gaia2023} which entitled us to perform a membership analysis.

\section{The membership analysis}
\label{membership}

An early proper motion analysis of NGC 6522 by \citet{Terndrup1998}, used the Bulge stars proper-motion survey of \citet{Spaenhauer1992}, identified within 2.5 arcminutes from the center, seven red giant branch (RGB) stars as cluster members, and employed them to estimate the cluster distance. Proper motions were also calculated by \citet{Rossi2015} employing images from two epochs in 1992 and 1995 from which they cleaned their CMD and estimated the distance. Taking advantage of the more accurate and precise proper motions presently available in the $Gaia$-DR3  catalogue, we are in position to revisit the membership question on NGC 6522 for a larger sample of stars, particularly for the variable stars in the field of the cluster.

As in previous works \citep[e.g.,][]{Arellano2023b}, we employed the method of \citet{Bustos2019} and high-quality astrometric data available in the $Gaia$-DR3, to separate likely cluster members from field stars. The method is based on two stages: 1) it aims to find groups of stars that possess similar characteristics in the four-dimensional space of the gnomonic coordinates ($X_{\rm t}$,$Y_{\rm t}$) and proper motions ($\mu_{\alpha*}$,$\mu_\delta$) by means of a clustering algorithm and 2) the analysis of the projected distribution of stars with different proper motions around of the mean proper motion of the cluster, in order to extract likely members that were missed in the first stage. Stars identified as members in the first stage are labeled in Table \ref{variables} as M1, whereas those identified after the second stage are labeled as M2. Stars with no available proper motions, hence with unknown membership status are label as UN. The specific details about this method are given in \citet{Bustos2019}.
The field of NGC 6522 is densely populated with more that 100 000 stars within a radius of 10 arcminutes. Since high densities are a problem for the present data releases $Gaia$ \citep{Fabricius2021, Luna2023}, more that 40\% of the stars have no mean proper motions and also there is no multi-band photometry for $G > 20$ mag, and for very few of those fainter than $G \sim 19.5$ mag.

\begin{figure*} 
\includegraphics[width=17.cm]{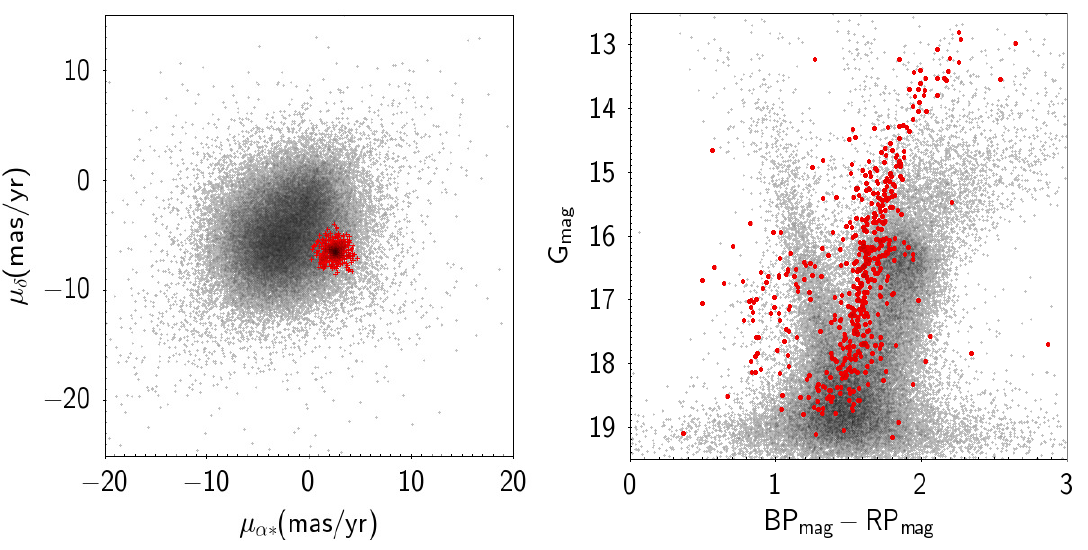}
\caption{VPD and CMD diagrams of $Gaia$-DR3 stars in the field of NGC 6522 (gray dots). The likely cluster member stars are displayed with red dots.}
    \label{Membership}
\end{figure*}

In Fig. \ref{Membership}, the vector point diagram (VPD) and CMD show the cluster identity. Only 689 likely members were found (red points in the figure), all of them within 2.5 arcmin of the cluster center. Being so few in such a dense field some contamination of these members cannot be ruled out. 

\subsection{The resulting OGLE CMD}
\begin{figure} 
\includegraphics[width=8.cm,height=8.cm]{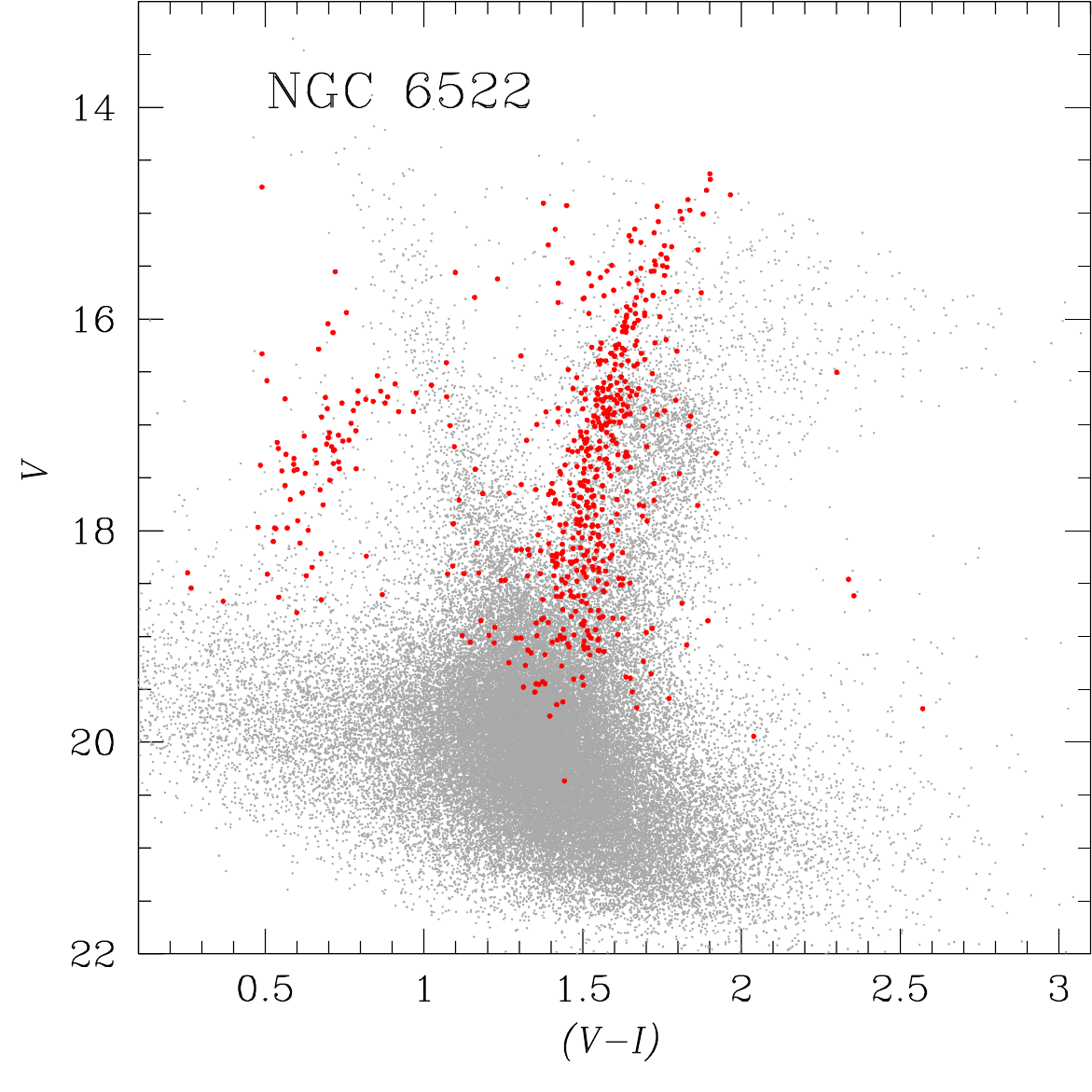}
\caption{OGLE \emph{VI} CMD diagram. The likely cluster member stars are displayed with red dots. Gray dots are nearly 60 000 field stars within 4 arcminutes from the cluster center.}
    \label{OGLECMD}
\end{figure}

Within a radius of about 4 arcminutes from the cluster center we found nearly 60 000 stars with \emph{VI} photometry in the OGLE IV collection. A \emph{VI} CMD diagram of this sample is shown in Fig. \ref{OGLECMD} (gray dots). A cross match with the $Gaia$-DR3 members identified in the previous section finds 580 likely member stars with \emph{VI} OGLE photometry. These are shown as red dots in the figure. Clear cluster sequences, like the RGB and the HB are seen. Let us now explore the cluster reddening and the positions in the CMD of the variable stars in the field of the cluster.

\section{Cluster reddening considerations}
\label{sec:reddening}

\begin{figure} 
\includegraphics[width=8.cm,height=8.cm]{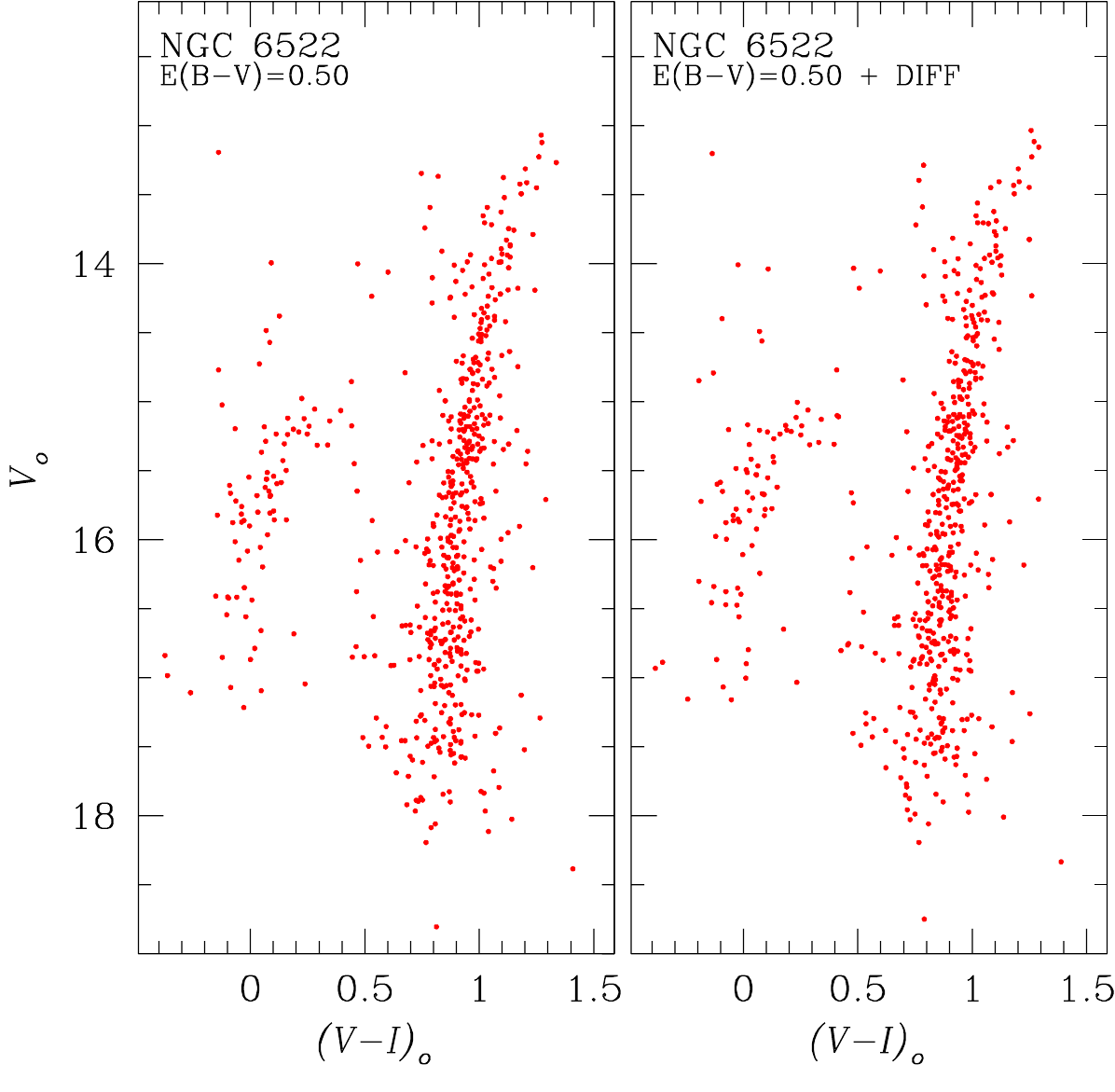}
\caption{Dereddened CMDs. In the left panel an overall constant $E(B-V)=0.50$ was applied. In the right panlel the excess $E(B-V)=0.50$ + DIFF was employed where  DIFF are the positional differential reddening values according to the map of \citet{AlonsoGaria2012}. The improvement by employing the differential values is marginal but evident.}
    \label{FIJODIFF}
\end{figure}

Typical of clusters in the Galactic bulge, NGC 6522 is subject not only to a large reddening ($E(B-V) \sim 0.50$ mag) but also to a differential one that we would like to consider before proceeding with our analysis. A differential reddening map for NGC 6522 is available in the work of \citet{AlonsoGaria2012} and we used it to correct the observed CMD. The map is a grid of 4809 differential reddening values within a field of 2.33 arcminutes from the cluster center. We applied the map to the likely members identified as described in $\S$ \ref{membership}. For each star we averaged
the differential reddening values for the four neighbouring values
in the grid and used the absolute
extinction zero-point of $E(B-V) = 0.50$ mag, according to the calibration  of \citet{Schlafly2011}, to obtain the reddening. Then, we corrected our $V$-band and $I$-band magnitudes for
each star by adopting a normal extinction $A_V = 3.12E(B-V)$ and the ratio $E(V-I)/E(B-V) = 1.259$. If a given member star is between 2.33 and 4 arcminutes from the center, it will not have a differential reddening associated; in these cases the average overall reddening of 0.50 mag was assigned. This is not exact and it may contribute to the noise in the differentially dereddened CMD. This was the case for 56 stars in our list of members. In Fig. \ref{FIJODIFF}, the CMDs dereddened with a constant $E(B-V)=0.50$ mag and the one differentially dereddened are compared. Although the difference between the two CMD's may seem marginal, there is a clear improvement after the differential reddening is considered, and we will use this CMD for the following discussions.

\begin{table*}
\footnotesize
\caption{Variable stars in the field of NGC 6522 within 4 arc minutes from the cluster center. See text in section \ref{Compilation} for explanation of the groupings. Stars labeled $Bl$ display clear Blazhko modulations.}
\label{variables}
\centering
\begin{tabular}{cccccccccc}
\hline
Variable & OGLE BLG & $Gaia$   & Variable   &  $P$   & HJD$_{\rm max}$     &  RA          & Dec.  &Memb       \\
Id  & Id& Source   & type   & (days)    
&  (days)    &  (J2000.0)   & (J2000.0) &status \\
\hline
\multicolumn{9}{c}{\bf CVSGC Variables }\\
\hline
V1  & RRLYR-12073 & 4050198351112412160 & RRc     & 0.269955   & 6694.8714 & 18:03:29.98 & -30:01:30.3 & M1 \\
V2 $Bl$ & RRLYR-12117 & 4050198316702394240 & RRab    & 0.474115   & 6175.6237 & 18:03:35.23 & -30:01:25.0 & M2 \\
V3  & RRLYR-12127 & 4050198110594112384 & RRc     & 0.288705   & 3522.9282 & 18:03:36.32 & -30:01:27.6 & M1 \\
V4 &     --      & 4050198110543863808 & RRab & 0.563832 &  4546.9025 & 18:03:37.15 & -30:01:56.7 & M1 \\
V5  & RRLYR-12154 & 4050198037650939648 & RRc     & 0.286826   & 2810.8836 & 18:03:40.25 & -30:02:47.7 & M1 \\
V6  & DSCT-06718  & 4050199519343205248 & DSCT    & 0.161273   & 7000.0709 & 18:03:42.58 & -30:01:34.7 & FS \\
V7  &     --      & 4050198381114239872 & LPV     & $\sim 700$ &     --    & 18:03:31.21 & -30:01:01.9 & FS \\
V8  & T2CEP-283   & 4050198312525412608 & CW      & 1.748029   & 4701.6116 & 18:03:33.63 & -30:01:15.1 & M1 \\
V9  & RRLYR-12099 & 4050198071876214784 & RRc     & 0.298717   & 7956.6875 & 18:03:33.63 & -30:03:09.3 & FS \\
V10 $Bl$ & RRLYR-12115 & 4050198110593995904 & RRab    & 0.557413   & 6850.5626 & 18:03:35.15 & -30:02:00.4 & M1/FS \\
V11 & RRLYR-12114 & 4050198110675567744 & RRab    & 0.615841   & 1294.8889 & 18:03:35.19 & -30:02:02.5 & M1/FS \\
V12 & RRLYR-12072 & 4050198278034844800 & RRc     & 0.256927   & 6386.9288 & 18:03:29.88 & -30:01:56.7 & FS \\
V13 & RRLYR-12108 & 4050198316752555520 & RRab    & 0.635979   & 3456.8979 & 18:03:34.28 & -30:01:45.0 & M1 \\
V14 & RRLYR-33606 & 4050198110543843456 & RRc     & 0.328105   & 5835.5691 & 18:03:35.40 & -30:01:55.3 & FS \\
V15 & T2CEP-284   & 4050198110543823616 & CW      & 1.797216   & 3937.4848 & 18:03:36.10 & -30:01:52.1 & M1 \\
\hline
\multicolumn{9}{c}{\bf OGs }\\
\hline
OG1 & RRLYR-11972 & 4050200859373592960 & RRab    & 0.427652   & 3426.8514 & 18:03:16.08 & -30:01:41.2 & FS \\
OG2 & RRLYR-11993 & 4050201237330707712 & RRab    & 0.454275   & 4681.6741 & 18:03:18.66 & -30:01:09.3 & FS \\
OG3 & RRLYR-12032 & 4050198209450683008 & RRab    & 0.440223   & 4278.6771 & 18:03:23.28 & -30:02:47.5 & FS \\
OG4 $Bl$& RRLYR-12044 & 4050197522140998912 & RRab    & 0.458734   & 7461.8947 & 18:03:24.17 & -30:05:13.6 & FS \\
OG5 & RRLYR-12185 & 4050197213017212544 & RRc     & 0.323838   & 2481.5494 & 18:03:45.27 & -30:03:40.7 & FS \\
OG6 & RRLYR-33568 & 4050197801306406656 & RRc     & 0.305943   & 7948.4722 & 18:03:18.77 & -30:04:01.4 & FS \\
\hline
\multicolumn{9}{c}{\bf Soszyński's LPVs }\\
\hline
V16 & LPV-193749  & 4050198174955503616 & SR     & 55.87      & 0699.5166 & 18:03:29.18 & -30:02:49.4 & M1 \\
V17 & LPV-194243  & 4050198106243226624 & OSARG   & 11.9158    &     --    & 18:03:34.59 & -30:02:00.3 & M1 \\
V18 & LPV-194274  & 4050198110543849472 & OSARG   & 23.597     &     --    & 18:03:34.85 & -30:02:11.5 & M1 \\
V19 & LPV-194290  & 4050198110594023296 & OSARG   & 14.922     &     --    & 18:03:34.99 & -30:01:56.3 & M1 \\
V20 & LPV-194311  & 4050198110543832064 & OSARG   & 51.18      & 1432.5462 & 18:03:35.20 & -30:02:10.7 & M1 \\
V21 & LPV-194316  & 4050198003164014080 & OSARG   & 16.948     &     --    & 18:03:35.27 & -30:03:02.6 & M1 \\
V22 & LPV-194544  & 4050198106236280704 & OSARG   & 17.827     &     --    & 18:03:37.92 & -30:02:07.4 & M1 \\
V23 & LPV-194566  & 4050199588012703488 & OSARG   & 12.131     &     --    & 18:03:38.19 & -30:00:51.7 & M1 \\
\hline
\multicolumn{9}{c}{\bf $Gaia$ member variables}\\
\hline
V24 & RRLYR-38961 & 4050198110543906816 & RRc     & 0.464360   & 7000.1644 & 18:03:36.13 & -30:01:48.7 & M1 \\
V25 &     --      & 4050198316752564352 & CW?     & 1.462408   & 7131.8794 & 18:03:34.55 & -30:01:38.0 & M2 \\

V26 &     --      & 4050198278034973312 & SR & --   & 7486.9229 & 18:03:32.57 & -30:01:57.6 & M2 \\
G1  &     --      & 4050199553702960256 & no var  &     --     &     --    & 18:03:42.72 & -30:01:21.0 & M1 \\
G2  &     --      & 4050198072010842624 & no var  &     --     &     --    & 18:03:35.60 & -30:02:40.1 & M1 \\
G3  &     --      & 4050197148521158144 & no var  &     --     &     --    & 18:03:34.51 & -30:04:18.2 & M1 \\
\hline
\end{tabular}
\end{table*}

\section{Variable stars in the field of NGC 6522}

\subsection{Compilation of variables data}
\label{Compilation}

In Table \ref{variables} we compile the variables in the field of NGC 6522 and organize them according to their source: the Catalogue of Variable Stars in Globular Clusters (CVSGC), OGLE (OGs), from \citet{Soszynski2013} and from $Gaia$-DR3. The different columns inform on the coordinates, id numbers, variable type, and ephemerides as well as the membership status assigned after our analysis.
The periods reported in Table \ref{variables} were calculated  by applying the
string-length method \citep{Burke1970, Dworetsky1983} as well as
{\tt period04} \citep{Lenz2005}, to the OGLE III and IV $I$-band data, which are the most abundant. 

\subsection{Listed in the Catalogue of Variable Stars in Globular Clusters }

In the CVSGC
\label{sec:CVSGC} \citep{cle01}, in its 2016 edition, there are listed 15 variables; five RRab, six RRc, two CW's or Pop II Cepheids, one high amplitude $\delta$ Scuti and one long period variable or LPV. Since the cluster is immersed in a very rich stellar field, only those variables within 2 arcminutes from the cluster center were included in the CVSGC table. Some of these may in fact be field stars (FS in Table\ref{variables}), e.g. the $\delta$ Scuti variable V6 \citep{McNamara2000}, and as we shall see below, the stars V7, V12 and V14 were also found to be field stars in our analysis. However, since variable members may indeed lie beyond 2 arc minutes from the center,  6 RR Lyrae between 2 and 4 arcminutes are listed in the CVSGC, which we have named OG1-6 for the purpose of the present work and list them in Table \ref{variables} in the sector named OG's. These six, genuine RR Lyrae stars were identified as field stars in our membership analysis.

\begin{figure*} 
\includegraphics[width=17cm]{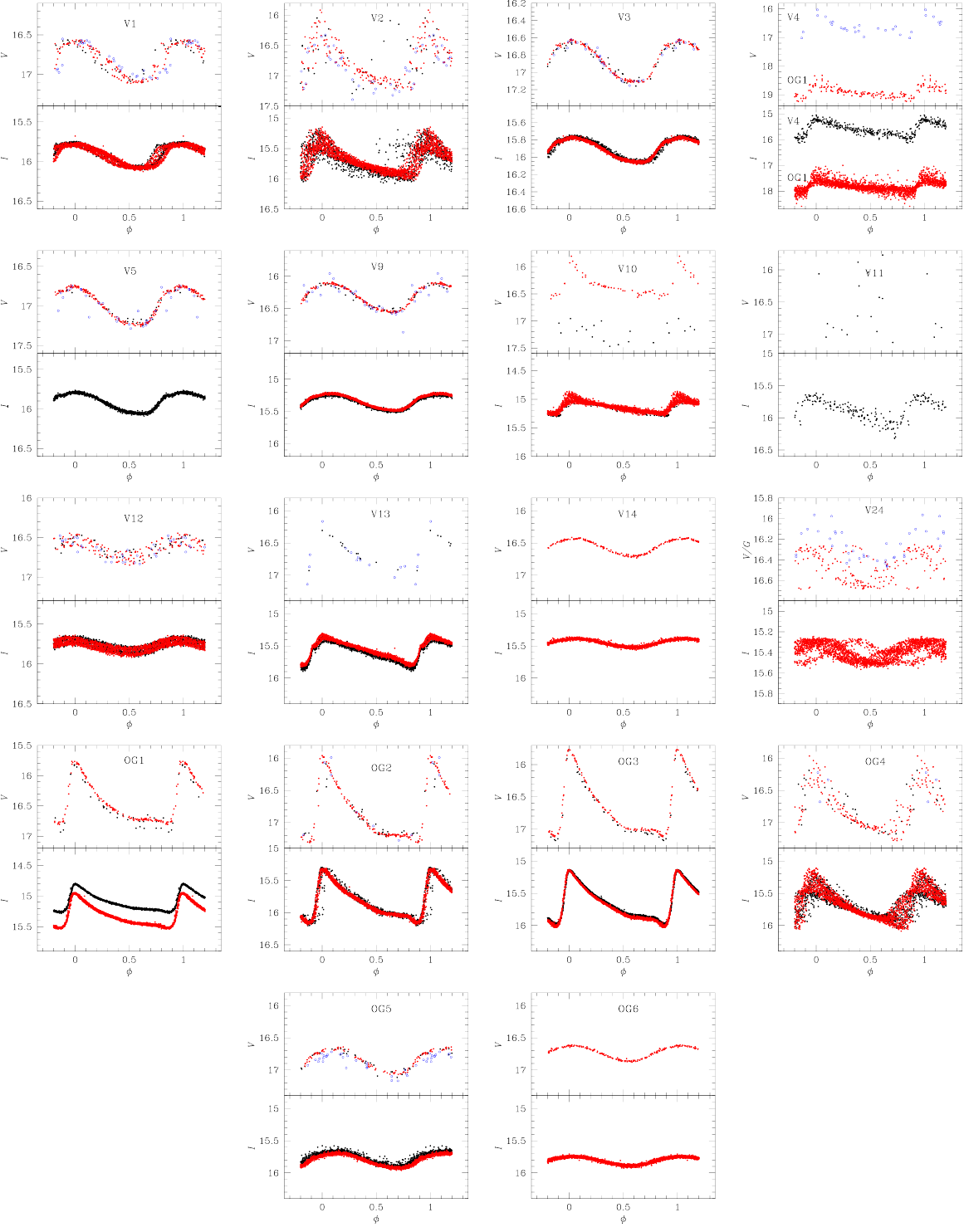}
\caption{Light curves of the RR Lyrae stars. Symbols are: black dots from OGLE III, red dots from OGLE IV and blue circles from $Gaia$-DR3 where $G$-band was converted into $V$-band \citep{Riello2021}, except for V24 as noted. For comments on peculiar stars see Appendix \ref{sec:IND_STARS}. We call special attention to the case of V4.}
   \label{RRLYR}
\end{figure*}

The OGLE and $Gaia$ light curves of the RR Lyrae and OG stars are displayed in Fig. \ref{RRLYR}.

\begin{figure*} 
\includegraphics[width=17.3cm]{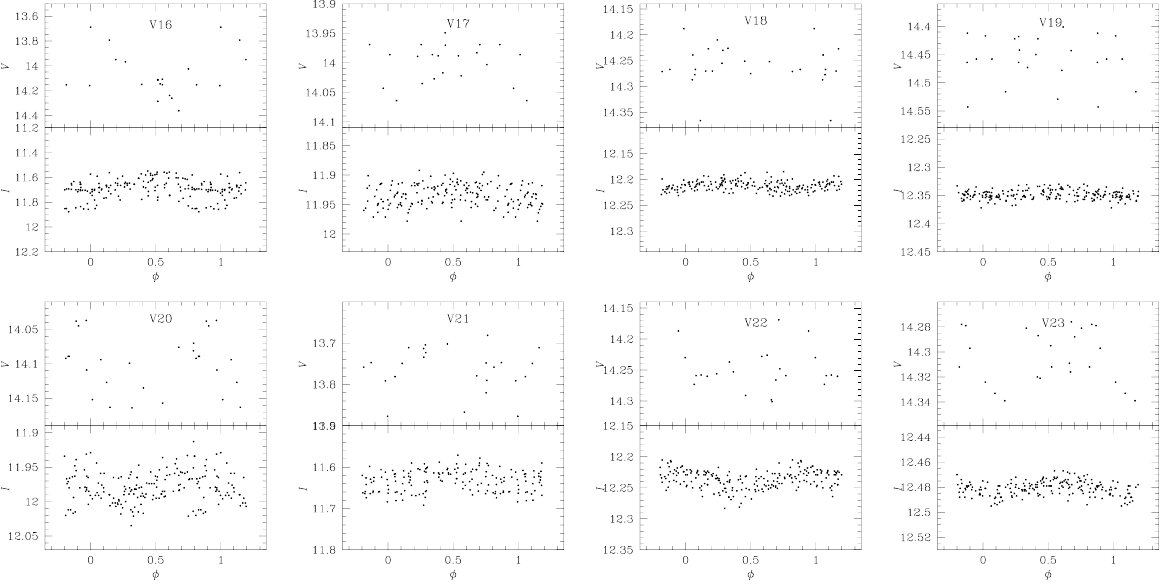}
\caption{OGLE light curves of 8 OSARG discovered by \citet{Soszynski2013}. We found them all to be cluster members, hence, variable names V16-V23 were assigned. Their periods are listed in Table \ref{variables}.}
    \label{OSs}
\end{figure*}

The CVSGC also informed us of 8 LPVs of small amplitude (or OGLE small amplitude red giants OSARGs) that were detected by \citet{Soszynski2013}.  These are included in Table \ref{variables} in the sector named Soszynski's LPVs. These eight OSARGs were found to be cluster members, hence we assigned them the variable names V16-V23, continuing with the numbering system in the CVSGC. Their light curves from the OGLE data base are displayed in Fig. \ref{OSs}.  

\subsection{OGLE and $Gaia$-DR3 databases}
\label{OGLE}

In the OGLE Catalogues, we found 177 variable stars with \emph{VI} light curves, within a radius of $\sim$ 4 arcminutes. The 29 stars listed in the previous subsection are included in this larger sample. 

\begin{figure*} 
\includegraphics[width=17.0cm]{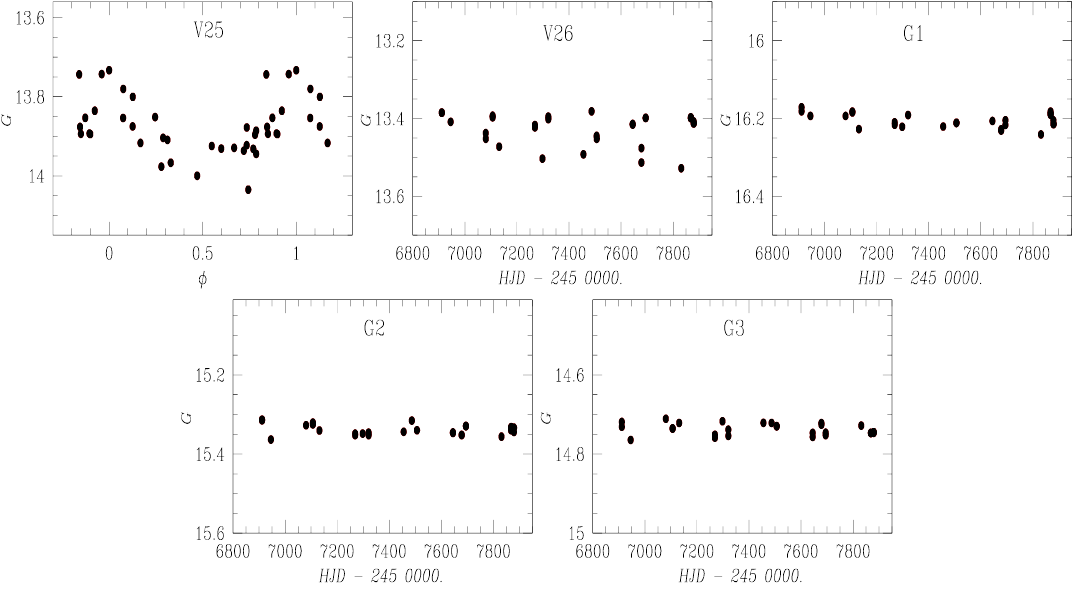}
\caption{$Gaia$ light curves five cluster members. G1-G3 turned out not to be conspicuous variables. V25 and V26 are in fact new member variables. Note that V25, with a period estimated, the light curve is phased whereas the others are plotted as a function of HJD.}
   \label{nuevas}
\end{figure*}

Similarly, within a 4 arcminutes radius from the cluster center, $Gaia$-DR3 reports 120 variables, with a few tens of epoch observations in the $Gaia$ photometric system.
Naturally there are cross matches between the OGLE and $Gaia$-DR3 variables. By a thorough inspection of these variables in both sources and their membership status resulting from the
$Gaia$-DR3 gnomonic coordinates and proper motions, we identified 6 new cluster member variables, and list them in the last sector of Table \ref{variables}, under the name "$Gaia$ member variables". Before assigning them a variable star identification, we inspected the $Gaia$ light curves to confirm or not  their variability. We found that the first three stars in this section of the table are genuine variables whose classifications will be discussed below. Therefore, we assign
them variable names V24, V25 and V26. The remainder three do not display a conspicuous variability within the time span of $Gaia$ observations, hence we refer to them as G1, G2 and G3. The light curves of these six stars in the $Gaia$ $G$-band are shown in Fig.  \ref{nuevas}. 

We stress at this point that among the stars classified as variables in the CVSGC, the 
stars V6, V7, V9, V12 and V14 are most likely field stars The star V4 is a cluster member RRab whose identification is worth a separate clarification note in the Appendix.

The NGC 6522 globular cluster variable star population now contains 21 member stars divided as follows: 5 RRab (V2, V4, V10, V11, V13), 4 RRc (V1, V3, V5, V24), 3 Pop II Cepheids or CWs (V8, V15, V25), 9 LPVs or SRs (V16-V23, V26).

Stars found to be field stars in our membership analysis, even if they are variables, are not further considered in the present work. Nevertheless, we note that there are 33 OGLE variables without a $Gaia$ counterpart and hence without proper motions.
Also, there are 31 OGLE variables with $Gaia$ counterpart but with no proper motion measurements. Thus, for these 64 stars in the 4 arcminutes radius field of the cluster we cannot say much on their cluster membership status. 
All of them are eclipsing binaries, and hence their position on the CMD does not offer a handle on the membership either. Their light curves are available in the OGLE Catalogues  \footnote{\url{http://ogle.astrouw.edu.pl}}.

\section{Variable stars in the CMD}
\label{sec:VAR_CMD}

\begin{figure*} 
\centering
\includegraphics[width=14.2cm]{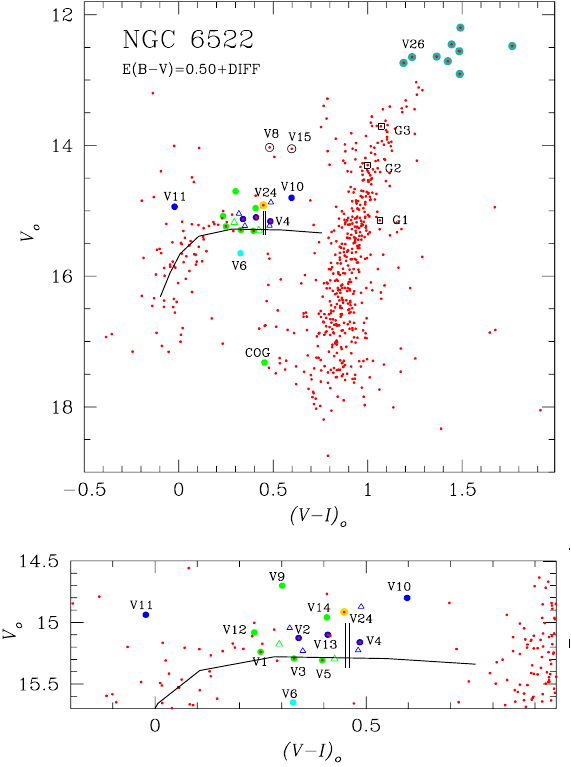}
\caption{CMD of cluster member stars in NGC 6522, diferentially dereddened. In colour symbols are all stars listed in Table \ref{variables} coded as follows: RRab stars (blue circles), RRc (green circles) CW (black open circles), LPV likely non members (open black squares), Soszynski’s member LPVs (darker green circles), non-member RRab and RRc (open triangles), $\delta$ Scuti V7 (turquoise circle) and V24 (yellow circle). All coloured star identifications having a red dot in the center are cluster member stars.  The bottom panel is an expansion of the HB region to facilitate the identification of stars. As a reference, the ZAHB for a central mass of 0.50 $M_{\odot}$, calculated by \citet{Yepez2022} using the Eggleton code \citep{Pols1997, Pols1998, KPS1997}. This locus was shifted to distance of 8.77 kpc.  The vertical black lines at the ZAHB mark the empirical red edge of the first overtone (FORE) instability strip \citep{Arellano2015,Arellano2016}. Its position with nearly all RRc stars to the left indicates that the employed average front extinction $E(B-V)=0.50$ is consistent with the pulsational modes border. RRab stars to the left of the FORE are siting in the bimodal region of the HB.}
   \label{CMD_HB}
\end{figure*}

The CMD of NGC 6522, differentially dereddened and displaying cluster member stars and those stars in Table \ref{variables} is shown in Fig. \ref{CMD_HB}.

\begin{figure*} 
\includegraphics[width=17.4cm]{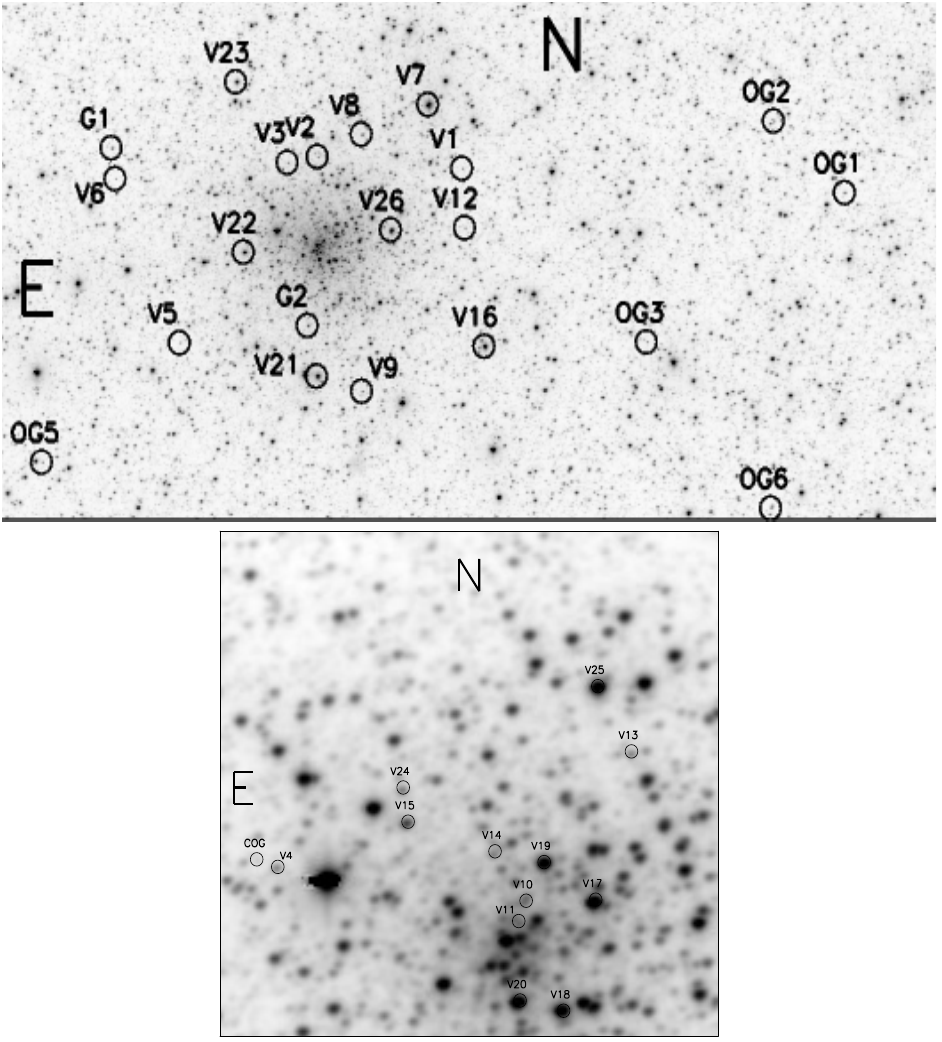}
\caption{NGC 6522 identification chart for the stars in Table \ref{variables} whose membership and identifications are discussed in the text. The bottom panel is an expansion of the core region. About the identification of V4 and its neighbour COG, see the note in the Appendix. The image employed for this chart is an OGLE $I$-band image. The upper field is about $7 \times 4$ arcmin$^2$ and lower field is about $1 \times 1$ arcmin$^2$.}
   \label{CHART}
\end{figure*}

An identification chart for the  stars in Table \ref{variables} is given in Fig. \ref{CHART}.

The positions of the variable stars in the CMD help confirming their type and complementing the proper motion information regarding their membership. However we must consider that, as we pointed out in section \ref{membership} related to the membership status of particular stars, since high
densities are a problem for the present $Gaia$ data releases, the resulting status may be in error. Hence the importance of the CMD diagram confrontation. Take for example the cases of the RRab stars V10 and V11. Their positions in the CMD suggest that these stars are not cluster members. However, their proper motions indicate they are cluster members of the 'M1' type. Nevertheless, being V10 and V11, immerse in the very core of the cluster, their photometry can easily be biased, thus, we cannot really assess their membership status.

On the other hand, some stars included in the CVSGC catalogue as cluster variables, are identified as field stars by our membership analysis as well as from their placement in the CMD; these are  V6, V9 and more marginally V12 and V14.

\begin{figure}[ht]
\includegraphics[width=8.0cm]{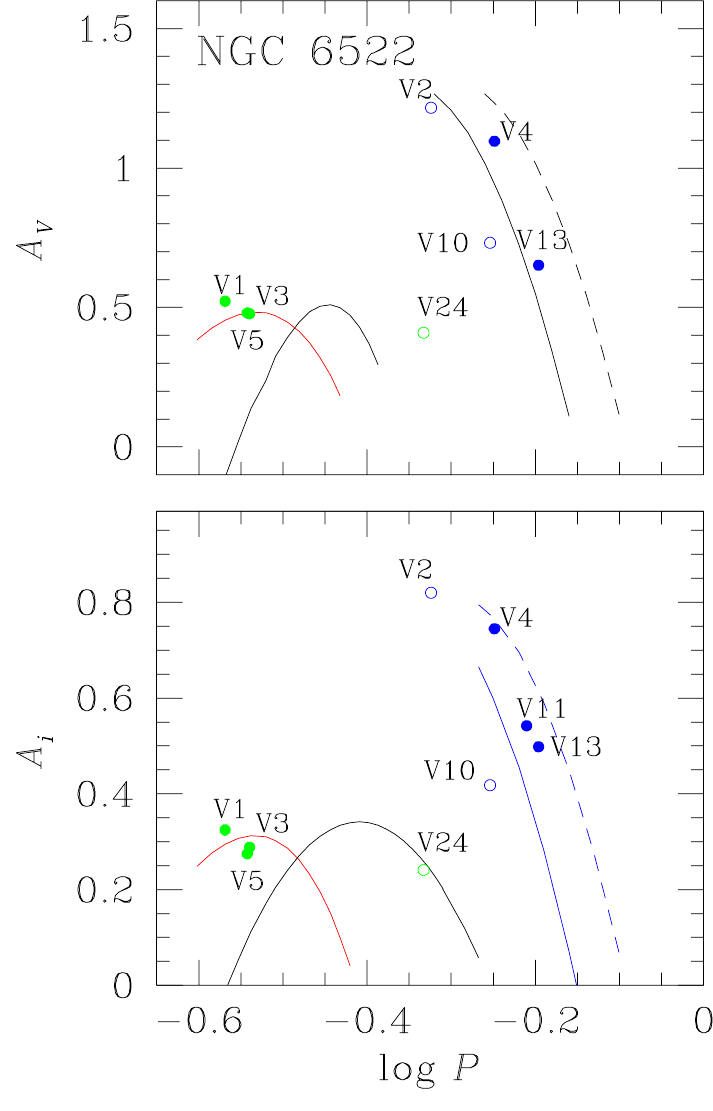}
\caption{Bailey's diagram of NGC6522. Vlue symbols are for RRab and RRc stars respectively. Open circles are for modulated stars. V24, the period modulating RRc stars has a peculiar position for this OoI type cluster. See section \ref{Oosterhoff} for details.}
   \label{Bailey}
\end{figure}

\section{[Fe/H] and M$_V$ from member RR Lyrae light curve Fourier decomposition}
\label{sec:RRLstars}

It is a well known fact that the Fourier decomposition of RR Lyrae light curves lead to accurate determinations of [Fe/H] and M$_V$, and hence, trough a knowledge of the interstellar reddening, to the cosmic distance of the star. A recent summary of the technique, the relevant calibrations, their zero points, and the results for a large number of RR Lyrae stars in a family of 39 globular clusters can be found in the paper by \citet{Arellano2023}.
The Fourier decomposition of a light curve involves fitting the curve with a series of sine or cosine harmonics of the form;

\begin{equation}
m(t) = A_0 ~+~ \sum_{k=1}^{N}{A_k ~\cos \left( \frac{2\pi}{P}k~(t-E) ~+~ \phi_k \right) },
\label{eq_foufit}
\end{equation}

\noindent
where $m(t)$ is the magnitude at time $t$, $P$ is the period and $E$ the epoch. A
linear
minimization routine is used to derive the amplitudes $A_k$ and phases $\phi_k$ of
each harmonic, from which the Fourier
parameters $\phi_{ij} = j\phi_{i} - i\phi_{j}$ and $R_{ij} = A_{i}/A_{j}$ are
calculated. The calculation of key physical parameters, like the atmospheric iron abundance, the luminosity and the stellar effective temperature, radius, and mass, is performed via well tested semi empirical calibrations and their zero points \citep{Arellano2023}.

To use RR Lyrae stars as indicators of the globular cluster metallicity and distance, we restrict to stars that have been identified as cluster members. In Table \ref{variables} we note three RRc members, V1, V3 and V5, and five RRab stars; V2, V4, V10, V11 and V13. The best $V$-band light curves for these stars come from the OGLE IV collection (red points in Fig. \ref{RRLYR}). Unfortunately the available $V$-band data of V4 are scarce for a proper Fourier decomposition and for V10 and V11, their membership status and/or photometry are dubious.

The mean magnitudes $A_0$, and the Fourier light curve fitting parameters of
individual RRab and RRc stars are listed in Table
~\ref{fourier_coeffs}. The corresponding individual and mean physical parameters are reported in Table  \ref{fisicos}.

\begin{table*}
\footnotesize
\caption{Fourier coefficients $A_{k}$ for $k=0,1,2,3,4$, and phases $\phi_{21}$,
$\phi_{31}$ and $\phi_{41}$, for RRab and RRc stars. The numbers in parentheses
indicate
the uncertainty on the last decimal places. Also listed is the deviation 
parameter $D_{\mbox{\scriptsize m}}$ (see Section~\ref{sec:RRLstars}).}
\centering                   
\begin{tabular}{lllllllllr}
\hline
Variable ID     & $A_{0}$    & $A_{1}$   & $A_{2}$   & $A_{3}$   & $A_{4}$   &
$\phi_{21}$ & $\phi_{31}$ & $\phi_{41}$ 
&  $D_{\mbox{\scriptsize m}}$ \\
     & ($V$ mag)  & ($V$ mag)  &  ($V$ mag) & ($V$ mag)& ($V$ mag) & & & & \\
\hline
\multicolumn{10}{c}{\bf RRab }\\
\hline
V2 & 16.826(13)  & 0.360(18) & 0.146(21) & 0.076(20) & 0.032(18) &3.975(172)& 8.195(307) &6.203(694) & 7.5\\
V4 &16.647 (3) & 0.227(4) & 0.172(5) & 0.161(4) & 0.050(4) & 3.249(391) & 7.513(521) & 5.431(900) & 13.8\\
V10 & 16.364(4) & 0.176(6) & 0.098(6) & 0.067(6) & 0.045(6) & 4.014(89) & 7.880(131) & 5.778(185) & 4.9\\
\hline
\multicolumn{10}{c}{\bf RRc }\\
\hline
V1 & 16.840(5)  & 0.255(7) & 0.034(7) & 0.009(7) & 0.009(7) &4.854(216)& 2.999(813) &
2.825(852) & --\\
V3 & 16.878(1) & 0.249(2) & 0.035(2) & 0.02(2) & 0.013(3) & 4.656(47) & 3.184(78) &
2.294(129) & --\\
V5 & 17.005(1) & 0.249(2) & 0.034(3) & 0.020(2) &0.012(2) & 4.465(68) & 3.188(105) &
2.072(178) & --\\

\hline
\end{tabular}
\label{fourier_coeffs}
\end{table*}

The iron abundance values are listed in two scales, that of \citet{Zinn1984}  [Fe/H]$_{\rm ZW}$, and in the spectroscopic scale of \citet{Carretta2009} [Fe/H]$_{\rm UVES}$
which are correlated by the equation [Fe/H]$_{\rm UVES}$ = $-0.413$ +0.130[Fe/H]$_{\rm ZW}-0.356$[Fe/H]$_{\rm ZW}^2$ (Carretta et al. 2009).

The absolute magnitudes $M_V$ were converted into distances by assuming individual differential reddenings calculated as explained in section \ref{sec:reddening}.

\noindent 
Let us recall that the calibration for [Fe/H] for RRab stars of Jurcsik
\& Kov\'acs (1996) is applicable to RRab stars with a {\it deviation parameter}
$D_m$,
defined by Jurcsik \& Kov\'acs (1996) and Kov\'acs \& Kanbur (1998), not exceeding an
upper limit. These authors suggest $D_m \leq 3.0$ for the light curves to be compatible with those of the stars used to set the calibration. Although occasionally this criterion is relaxed and $D_m \sim 5.0$, in the present case of neither V2, V4 nor V10 meeting the criterion and display very noisy light curves, thus we opted for not considering them as physical parameters indicators.

In conclusion, our best estimates of a mean metallicity and distance of NGC 6522 via the Fourier light curve decomposition, come from its three member RRc stars V1, V3, V5, with reliable OGLE $V$-band data, from which we find the averages [Fe/H]$_{\rm ZW}$=--1.28 $\pm$0.09; [Fe/H]$_{\rm UVES}$=--1.16 $\pm$0.09; and $D=8.77  \pm 0.16$ kpc. At this point it is convenient to note the comparison of these results with the high resolution spectroscopic value of [Fe/H]$=-1.0 \pm 0.2$ obtained by \citet{Barbuy2009} via the analysis of 8 giant stars in NGC 6522, and with the distance $8.120 \pm 0.929$ kpc from $Gaia$-EDR3 data analysis performed by \citet{Baumgardt2021} respectively (see section \ref {sec:distance} for further comments on the distance).

\begin{table*}
\footnotesize
\caption{\small Physical parameters for the RRab and RRc stars. The numbers in
parentheses indicate the uncertainty on the last 
decimal place.}
\centering
\label{fisicos}
\hspace{0.01cm}
 \begin{tabular}{lcccccccc}
\hline 
Star&[Fe/H]$_{\rm ZW}$  &[Fe/H]$_{\rm UVES}$ &$M_V$ & D(kpc)&log~$T_{\rm eff}$  & log$(L/{\rm
L_{\odot}})$ &
$M/{\rm M_{\odot}}$&$R/{\rm R_{\odot}}$\\
\hline
\multicolumn{9}{c}{\bf RRab }\\
\hline
V2 &$-1.17(30)^a$& $-1.06(24)^a$& 0.66(3) &8.32(11)& 3.821(78) & 1.636(11) & 0.70(9) & 5.03(9) \\
V4 &$-2.1(49)^a$& $-2.34(76)^a$& 0.75(1) &7.31(2)& 3.797(112) & 1.634(2) & 0.72(9) & 5.61(2) \\
V10 &$-1.78(12)^a$& $-1.78(16)^a$& 0.74(1) &6.65(3) &3.802(22) & 1.624(3) & 0.68(18) & 5.43(2) \\
\hline
Weighted &&&&&&\\
Mean & $-1.71(49)^a$& $-1.58(13)^a$ & 0.74(6) &7.10(0.93) &3.803(1) & 1.630(1) & 0.68(17) & 5.53(1)\\
\hline 
\multicolumn{9}{c}{\bf RRc }\\
\hline
V1 &$-1.15(13) $& $-1.03(13) $& 0.619(33) & 8.54(13)&3.873(5) & 1.652(13) & 0.60(3) & 4.05(6) \\
V3 &$-1.29(14) $ & $-1.17(13) $& 0.591(9) &8.78(4)& 3.858(1) & 1.663(1) & 0.66(1) & 4.38(2) \\
V5 &$-1.26(18) $ & $-1.14(17) $& 0.603(1) & 8.77(4)&3.847(1) & 1.659(4) & 0.76(1) & 4.59(2) \\
\hline
Weighted &&&&&&\\
Mean & $-1.28(9)$& $-1.16(9)$ & 0.598(10) & 8.77(16)&3.854(1) & 1.661(1) & 0.70(1) & 4.46(1)\\
\hline 
\end{tabular}
\center{$a$: Non of the light curves of these RRab stars fulfill the $D_m$ parameter condition, thus thier metallicity values are spurious.\\}

\end{table*}

\subsection{On the Oosterhoff type of NGC 6522}
\label{Oosterhoff}

The average of the periods of the four RRab member stars is 0.57 days. This and the rather high iron abundance of  [Fe/H]$_{\rm ZW}$=--1.28, clearly identify NGC 6522 as an Oo I type cluster.

A Period-Amplitud, or Bailey's diagram for the RR Lyrae stars in NGC 6522 in the $V$- and $I$-bands is shown in Fig. \ref{Bailey}. The loci 
in the top panel and to the right, are for RRab stars (unevolved
continuous and evolved segmented) in M3 according to \citep{Cacciari2005}. \citet{Kunder2013a}  found the black parabola for the RRc stars from
14 OoII clusters and \citet{Arellano2015}  calculated the red parabolas
from a sample of RRc stars in five OoI clusters, excluding variables with
Blazhko. In the bottom panel, the continuous and segmented blue
lines were constructed by \citet{Kunder2013b}. The black parabola was
calculated by \citet{Yepez20}, using 28 RRc stars from seven OoII
clusters.

The stars distribution clearly reflect the OoI nature of NGC 6522, as indicated by the RRab average period. The position of V4 corroborates our identification. The peculiar period modulated V24 does not follow the trend the non-modulated stars for an OoI cluster (red parabola).

\section{Comments on the Cluster Distance}
\label{sec:distance}

Our first estimation of cluster distance comes from the Fourier decomposition and the absolute magnitude calibration for the 3 RRc members stars V1, V3 and V5, yielding an average distance of $8.77\pm0.16$ kpc. 

A second independent estimate can be made by employing the $I$-band RR Lyrae P-L relation derived by \citet{Catelan2004};

 \begin{equation}
M_I = 0.471 - 1.132 {\rm log} P + 0.205 {\rm log} Z,
\label{eqn:PL_RRI}
\end{equation}

\noindent
with ${\rm log} Z =[M/H] -1.765$. We applied these equations to 4 RRab and 3
RRc cluster member stars in 
Table \ref{fisicos}.
The periods for the RRc stars were fundamentalized following the
period ratio $P_1/P_0 =0.7454$ in double mode stars \citet{Catelan2009}. We employed the average overall reddening $E(B-V)=0.50$ (see section \ref{sec:reddening}). The resulting mean distance was 8.46$\pm$0.80 kpc, in very good agreement with the Fourier approach above.

In their paper with accurate Galactic globular cluster distance determinations using $Gaia$-EDR3 and Hubble Space Telescope data, and distances compilation from the literature, \citet{Baumgardt2021} note 9 independent distance values for NGC 6522 obtained between 1998 and 2021. These distances range between 6.1 kpc(the oldest) and 8.9 kpc (the newest); (see also the "Fundamental parameters of Galactic globular clusters" compilation by  Baumgardt et al.  (2023)\footnote{\url{https://people.smp.uq.edu.au/HolgerBaumgardt/globular}}. These authors tend to prefer the lower determinations and opted for an average of $7.29 \pm 0.20$ kpc. 
It has been demonstrated \citep{Arellano2023} that the globular cluster distances found from the Fourier decomposition of RR Lyrae stars agree within less than 2 kpc with the distances of \citet{Baumgardt2021}, in a range of 40 kpc. While this is true for the comparison of NGC 6522, we should remark that our results prefer the larger distance values, in agreement with their $Gaia$-EDR3 results ranging $8.1-8.9$ kpc.

\section{The case of the V24: A period modulating star}

The light curve of V24 in the $I$-band, (Fig. \ref{RRLYR}), is dense and displays clear phase (hence period) modulations and no amplitude modulations are apparent, as would be the case in a regular double-mode star. We opted for analysing the times of maximum $I$-band light since the OGLE available data span 2674 days, which considering the short pulsating period, is long enough to study the modulations behaviour. A thorough analysis of the light curve phased with the ephemerides in Table \ref{variables}, enable us to identify 66 times of maximum light in the about 4000 days of observations by OGLE IV. To explore the behaviour of the residuals between the observed and the calculated maxima given a fixed ephemerides, $O-C$ for short, we employed the approach discussed in detail in the the papers by \citet{Arellano2016a,Arellano2018}. In brief, to
predict the time of maximum $C$ one adopts an ephemeris of the form $C=E_0 +P_1 N$; where $E_0$ is an adopted origin or epoch, $P_1$ is the period at $E_0$ and
$N$ is the number of cycles elapsed between $E_0$ and $C$. An estimate of the number of cycles between the observed time of maximum $O$ and the reference $E_0$ is simply, $N= \lfloor (O - E_0)/P_1 \rfloor$ where the incomplete brackets indicate the rounding down to the
nearest integer.

\begin{figure}[ht]
\includegraphics[width=8.0cm]{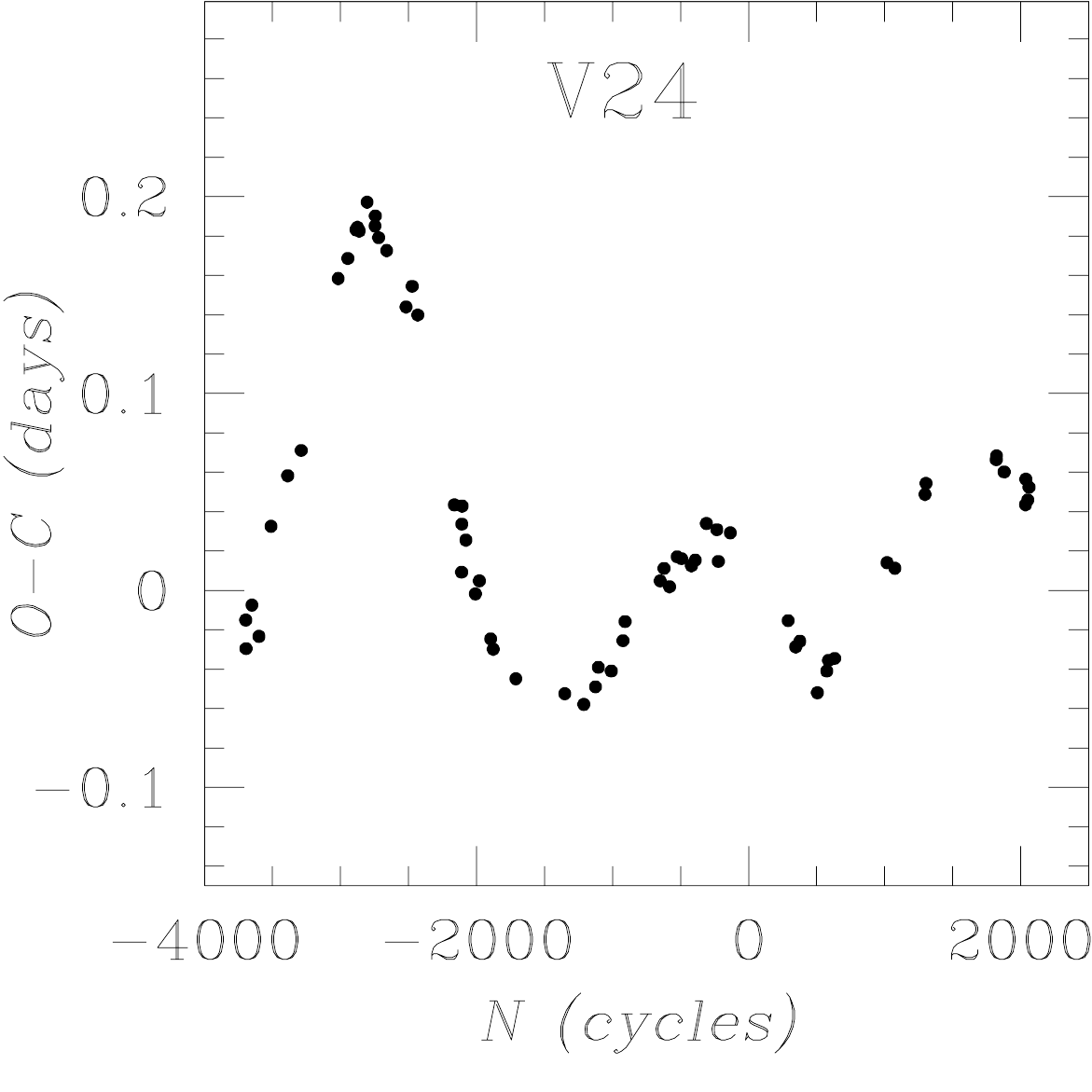}
\caption{$O-C$ diagram of the V24, built with the ephemerides $C=0.46436 N + 2457000.1614$ days, where $N$ is the number of cycles.}
   \label{OmC}
\end{figure}

The resulting $O-C$ diagram of V24 is shown in Fig. \ref{OmC} and displays well developed modulations with a separation between maxima of about 2560 cycles or about 1089 days. Whether these modulations are periodic is difficult to say with the present data. Phase and amplitude modulations in RRc stars in the OGLE IV data base have been studied by \citet{Netzel2018}. One difference between V24 and most of the stars exemplified by Netzel and collaborators is that in V24 prewhitening of the overtone frequency, $f_1=1/P_1$, does not fully remove the signal around that frequency and substantial signal remains even after several attempts.

\section{Summary and conclusions}
\label{summary}

In the present paper we have employed the OGLE III, OGLE IV and $Gaia$-DR3 data to address several conclusions. From the membership analysis of about 100 000 stars in a field of 10 arcminutes around the cluster and using the gnomic positions and proper motions in the $Gaia$-DR3 facility, we identified 689 likely cluster member stars. Out of these, only 580 stars had \emph{VI} photometry within the 60 000 stars with such photometry in the OGLE data base. Nevertheless, these stars conform a clean CMD diagram with clear evolutionary regions such as the HB and the RGB.

We corrected the CMD from differential reddening using the reddening map of \citet{AlonsoGaria2012}, and noticed a mild but clear improvement over the correction with an overall average reddening.

The final membership status of all variable stars under discussion was assigned by comparing the results from the membership analysis and the star position in the CMD. This led to the conclusion that stars V6, V7, V9, V12 and V14, in the list of the CVSGC are not cluster members. For V4 we conclude, after the discussion on its identification (see notes in the Appendix) that it is a cluster member RRab. Not a clear conclusion could be reached for V10 and V11 since their position in the core of the cluster make their photometry dubious. 

On the other hand we confirm the 8 OSARGs identified by \citet{Soszynski2013} to be cluster members and assigned them a "V" variable star name following the CVSGC.

From a the large number of variables in the field of NGC 6522 reported by both OGLE and $Gaia$-DR3, we identified six new cluster members. However for three of them we could not detect any variation within the time span of the $Gaia$ observations. Hence, we assigned variable names to the three member variables V24-V26, and retain the other four non-variable members as G1-G3. Among the four member variables, V24 was classified as RRc ongoing very strong and irregular period modulations, V25 as probably heavily reddened CW and V26 as a semiregular or SR giant star.

In conclusion, the variable star population of NGC 6522 (including for the time being V10 and V11) now contains 20 member stars grouped as 4 RRab, 4 RRc, 3 CWs, and 9 LPVs.

The Fourier light curve decomposition of the RRc stars with OGLE reliable $V$-band data V1, V3 and V5, lead to the mean metallicity and distance [Fe/H]$_{\rm ZW}$=-1.28 $\pm$0.09; [Fe/H]$_{\rm UVES}$=-1.16 $\pm$0.09; and $D=8.77  \pm 0.16$ kpc. The $I$-band P-L relation of \citet{Catelan2004} conducts to a distance of 8.46$\pm$0.80 kpc.
These distance determinations are in excellent agreement with the distance based on $Gaia$-EDR3 results of $8.1-8.9$ kpc given in  \citet{Baumgardt2021}.

Finally the newly identified member variable V24 is classified as period-modulating RRc star for which no amplitude modulations are detected. An $O-C$ analysis displays an irregular period modulation with variable amplitude and a time scale of about 1100 days. These long-term period modulations are not the result of a  Blazhko-like nature and still await a theoretical explanation.

\section*{Acknowledgments}
This publication was made possible by grant
IG 100620 from the DGAPA-UNAM (Mexico).
We have made an extensive use of the SIMBAD and ADS services, for which we
are thankful.

\bibliographystyle{spr-mp-nameyear-cnd}
\bibliography{bibliografia}

\begin{thebibliography}{44}
\ifx \bisbn   \undefined \def \bisbn  #1{ISBN #1}\fi
\ifx \binits  \undefined \def \binits#1{#1} \fi
\ifx \bauthor  \undefined \def \bauthor#1{#1} \fi
\ifx \batitle  \undefined \def \batitle#1{#1} \fi
\ifx \bjtitle  \undefined \def \bjtitle#1{#1}\fi
\ifx \bvolume  \undefined \def \bvolume#1{\textbf{#1}}\fi
\ifx \byear  \undefined \def \byear#1{#1} \fi
\ifx \bissue  \undefined \def \bissue#1{#1} \fi
\ifx \bfpage  \undefined \def \bfpage#1{#1} \fi
\ifx \blpage  \undefined \def \blpage #1{#1} \fi
\ifx \burl  \undefined \def \burl#1{\textsf{#1}} \fi
\ifx \doiurl  \undefined \def \doiurl#1{\textsf{#1}} \fi
\ifx \betal  \undefined \def \betal{\textit{et al.}} \fi
\ifx \binstitute  \undefined \def \binstitute#1{#1} \fi
\ifx \binstitutionaled  \undefined \def \binstitutionaled#1{#1} \fi
\ifx \bctitle  \undefined \def \bctitle#1{#1} \fi
\ifx \beditor  \undefined \def \beditor#1{#1} \fi
\ifx \bpublisher  \undefined \def \bpublisher#1{#1} \fi
\ifx \bbtitle  \undefined \def \bbtitle#1{#1} \fi
\ifx \bedition  \undefined \def \bedition#1{#1} \fi
\ifx \bseriesno  \undefined \def \bseriesno#1{#1} \fi
\ifx \blocation  \undefined \def \blocation#1{#1} \fi
\ifx \bsertitle  \undefined \def \bsertitle#1{#1} \fi
\ifx \bsnm \undefined \def \bsnm#1{#1} \fi
\ifx \bsuffix \undefined \def \bsuffix#1{#1} \fi
\ifx \bparticle \undefined \def \bparticle#1{#1} \fi
\ifx \barticle \undefined \def \barticle#1{#1} \fi
\ifx \bconfdate \undefined \def \bconfdate #1{#1} \fi
\ifx \botherref \undefined \def \botherref #1{#1} \fi
\ifx \url \undefined \def \url#1{\textsf{#1}} \fi
\ifx \bchapter \undefined \def \bchapter#1{#1} \fi
\ifx \bbook \undefined \def \bbook#1{#1} \fi
\ifx \bcomment \undefined \def \bcomment#1{#1} \fi
\ifx \oauthor \undefined \def \oauthor#1{#1} \fi
\ifx \citeauthoryear \undefined \def \citeauthoryear#1{#1} \fi
\ifx \endbibitem  \undefined \def \endbibitem {}\fi
\ifx \bconflocation  \undefined \def \bconflocation#1{#1} \fi
\ifx \arxivurl  \undefined \def \arxivurl#1{\textsf{#1}} \fi

\bibitem[\protect\citeauthoryear{{Alonso-Garc{\'\i}a}
  et~al.}{2012}]{AlonsoGaria2012}
\begin{barticle}
\bauthor{\bsnm{{Alonso-Garc{\'\i}a}}, \binits{J.}},
\bauthor{\bsnm{{Mateo}}, \binits{M.}},
\bauthor{\bsnm{{Sen}}, \binits{B.}},
\bauthor{\bsnm{{Banerjee}}, \binits{M.}},
\bauthor{\bsnm{{Catelan}}, \binits{M.}},
\bauthor{\bsnm{{Minniti}}, \binits{D.}},
\bauthor{\bsnm{{von Braun}}, \binits{K.}}:
\bjtitle{\aj}
\bvolume{143}(\bissue{3}),
\bfpage{70}
(\byear{2012})
\end{barticle}
\endbibitem

\bibitem[\protect\citeauthoryear{{Arellano Ferro} et~al.}{2015}]{Arellano2015}
\begin{barticle}
\bauthor{\bsnm{{Arellano Ferro}}, \binits{A.}},
\bauthor{\bsnm{{Mancera Pi{\~n}a}}, \binits{P.E.}},
\bauthor{\bsnm{{Bramich}}, \binits{D.M.}},
\bauthor{\bsnm{{Giridhar}}, \binits{S.}},
\bauthor{\bsnm{{Ahumada}}, \binits{J.A.}},
\bauthor{\bsnm{{Kains}}, \binits{N.}},
\bauthor{\bsnm{{Kuppuswamy}}, \binits{K.}}:
\bjtitle{\mnras}
\bvolume{452}(\bissue{1}),
\bfpage{727}
(\byear{2015})
\end{barticle}
\endbibitem

\bibitem[\protect\citeauthoryear{{Arellano Ferro} et~al.}{2016a}]{Arellano2016}
\begin{barticle}
\bauthor{\bsnm{{Arellano Ferro}}, \binits{A.}},
\bauthor{\bsnm{{Luna}}, \binits{A.}},
\bauthor{\bsnm{{Bramich}}, \binits{D.M.}},
\bauthor{\bsnm{{Giridhar}}, \binits{S.}},
\bauthor{\bsnm{{Ahumada}}, \binits{J.A.}},
\bauthor{\bsnm{{Muneer}}, \binits{S.}}:
\bjtitle{\apss}
\bvolume{361}(\bissue{5}),
\bfpage{175}
(\byear{2016}a)
\end{barticle}
\endbibitem

\bibitem[\protect\citeauthoryear{{Arellano Ferro}
  et~al.}{2016b}]{Arellano2016a}
\begin{barticle}
\bauthor{\bsnm{{Arellano Ferro}}, \binits{A.}},
\bauthor{\bsnm{{Ahumada}}, \binits{J.A.}},
\bauthor{\bsnm{{Kains}}, \binits{N.}},
\bauthor{\bsnm{{Luna}}, \binits{A.}}:
\bjtitle{\mnras}
\bvolume{461}(\bissue{1}),
\bfpage{1032}
(\byear{2016}b)
\end{barticle}
\endbibitem

\bibitem[\protect\citeauthoryear{{Arellano Ferro} et~al.}{2018}]{Arellano2018}
\begin{barticle}
\bauthor{\bsnm{{Arellano Ferro}}, \binits{A.}},
\bauthor{\bsnm{{Rosenzweig}}, \binits{P.}},
\bauthor{\bsnm{{Luna}}, \binits{A.}},
\bauthor{\bsnm{{Deras}}, \binits{D.}},
\bauthor{\bsnm{{Muneer}}, \binits{S.}},
\bauthor{\bsnm{{Giridhar}}, \binits{S.}},
\bauthor{\bsnm{{Michel}}, \binits{R.}}:
\bjtitle{Astronomische Nachrichten}
\bvolume{339},
\bfpage{158}
(\byear{2018})
\end{barticle}
\endbibitem

\bibitem[\protect\citeauthoryear{{Arellano Ferro} et~al.}{2023}]{Arellano2023b}
\begin{barticle}
\bauthor{\bsnm{{Arellano Ferro}}, \binits{A.}},
\bauthor{\bsnm{{Rojas Galindo}}, \binits{F.C.}},
\bauthor{\bsnm{{Bustos Fierro}}, \binits{I.H.}},
\bauthor{\bsnm{{Muneer}}, \binits{S.}},
\bauthor{\bsnm{{Yepez}}, \binits{M.A.}},
\bauthor{\bsnm{{Giridhar}}, \binits{S.}}:
\bjtitle{\mnras}
\bvolume{519}(\bissue{2}),
\bfpage{2451}
(\byear{2023})
\end{barticle}
\endbibitem

\bibitem[\protect\citeauthoryear{{Arellano Ferro}}{2023}]{Arellano2023}
\begin{botherref}
\oauthor{\bsnm{{Arellano Ferro}}, \binits{A.}}:
To appear in IAU Symp No. 376: At the cross-roads of astrophysics and
  cosmology: Period luminosity relations in the 2020s. (arXiv:2306.01175),
2306
(2023)
\end{botherref}
\endbibitem

\bibitem[\protect\citeauthoryear{{Baade}}{1946}]{Baade1946}
\begin{barticle}
\bauthor{\bsnm{{Baade}}, \binits{W.}}:
\bjtitle{\pasp}
\bvolume{58}(\bissue{343}),
\bfpage{249}
(\byear{1946})
\end{barticle}
\endbibitem

\bibitem[\protect\citeauthoryear{{Barbuy} et~al.}{2009}]{Barbuy2009}
\begin{barticle}
\bauthor{\bsnm{{Barbuy}}, \binits{B.}},
\bauthor{\bsnm{{Zoccali}}, \binits{M.}},
\bauthor{\bsnm{{Ortolani}}, \binits{S.}},
\bauthor{\bsnm{{Hill}}, \binits{V.}},
\bauthor{\bsnm{{Minniti}}, \binits{D.}},
\bauthor{\bsnm{{Bica}}, \binits{E.}},
\bauthor{\bsnm{{Renzini}}, \binits{A.}},
\bauthor{\bsnm{{G{\'o}mez}}, \binits{A.}}:
\bjtitle{\aap}
\bvolume{507}(\bissue{1}),
\bfpage{405}
(\byear{2009})
\end{barticle}
\endbibitem

\bibitem[\protect\citeauthoryear{{Baumgardt} and
  {Vasiliev}}{2021}]{Baumgardt2021}
\begin{barticle}
\bauthor{\bsnm{{Baumgardt}}, \binits{H.}},
\bauthor{\bsnm{{Vasiliev}}, \binits{E.}}:
\bjtitle{\mnras}
\bvolume{505}(\bissue{4}),
\bfpage{5957}
(\byear{2021})
\end{barticle}
\endbibitem

\bibitem[\protect\citeauthoryear{{Bica} et~al.}{2019}]{Bica2019}
\begin{barticle}
\bauthor{\bsnm{{Bica}}, \binits{E.}},
\bauthor{\bsnm{{Pavani}}, \binits{D.B.}},
\bauthor{\bsnm{{Bonatto}}, \binits{C.J.}},
\bauthor{\bsnm{{Lima}}, \binits{E.F.}}:
\bjtitle{\aj}
\bvolume{157}(\bissue{1}),
\bfpage{12}
(\byear{2019}).
\arxivurl{1812.10292}.
doi:\doiurl{10.3847/1538-3881/aaef8d}
\end{barticle}
\endbibitem

\bibitem[\protect\citeauthoryear{{Burke} et~al.}{1970}]{Burke1970}
\begin{barticle}
\bauthor{\bsnm{{Burke}}, \binits{J.} \bsuffix{Edward~W.}},
\bauthor{\bsnm{{Rolland}}, \binits{W.W.}},
\bauthor{\bsnm{{Boy}}, \binits{W.R.}}:
\bjtitle{\jrasc}
\bvolume{64},
\bfpage{353}
(\byear{1970})
\end{barticle}
\endbibitem

\bibitem[\protect\citeauthoryear{{Bustos Fierro} and
  {Calder{\'o}n}}{2019}]{Bustos2019}
\begin{barticle}
\bauthor{\bsnm{{Bustos Fierro}}, \binits{I.H.}},
\bauthor{\bsnm{{Calder{\'o}n}}, \binits{J.H.}}:
\bjtitle{\mnras}
\bvolume{488},
\bfpage{3024}
(\byear{2019})
\end{barticle}
\endbibitem

\bibitem[\protect\citeauthoryear{{Cacciari} et~al.}{2005}]{Cacciari2005}
\begin{barticle}
\bauthor{\bsnm{{Cacciari}}, \binits{C.}},
\bauthor{\bsnm{{Corwin}}, \binits{T.M.}},
\bauthor{\bsnm{{Carney}}, \binits{B.W.}}:
\bjtitle{\aj}
\bvolume{129}(\bissue{1}),
\bfpage{267}
(\byear{2005})
\end{barticle}
\endbibitem

\bibitem[\protect\citeauthoryear{{Carretta} et~al.}{2009}]{Carretta2009}
\begin{barticle}
\bauthor{\bsnm{{Carretta}}, \binits{E.}},
\bauthor{\bsnm{{Bragaglia}}, \binits{A.}},
\bauthor{\bsnm{{Gratton}}, \binits{R.}},
\bauthor{\bsnm{{D'Orazi}}, \binits{V.}},
\bauthor{\bsnm{{Lucatello}}, \binits{S.}}:
\bjtitle{\aap}
\bvolume{508}(\bissue{2}),
\bfpage{695}
(\byear{2009})
\end{barticle}
\endbibitem

\bibitem[\protect\citeauthoryear{{Catelan}}{2009}]{Catelan2009}
\begin{barticle}
\bauthor{\bsnm{{Catelan}}, \binits{M.}}:
\bjtitle{\apss}
\bvolume{320},
\bfpage{261}
(\byear{2009})
\end{barticle}
\endbibitem

\bibitem[\protect\citeauthoryear{{Catelan} et~al.}{2004}]{Catelan2004}
\begin{barticle}
\bauthor{\bsnm{{Catelan}}, \binits{M.}},
\bauthor{\bsnm{{Pritzl}}, \binits{B.J.}},
\bauthor{\bsnm{{Smith}}, \binits{H.A.}}:
\bjtitle{\apjs}
\bvolume{154},
\bfpage{633}
(\byear{2004})
\end{barticle}
\endbibitem

\bibitem[\protect\citeauthoryear{{Clement} et~al.}{2001}]{cle01}
\begin{barticle}
\bauthor{\bsnm{{Clement}}, \binits{C.M.}},
\bauthor{\bsnm{{Muzzin}}, \binits{A.}},
\bauthor{\bsnm{{Dufton}}, \binits{Q.}},
\bauthor{\bsnm{{Ponnampalam}}, \binits{T.}},
\bauthor{\bsnm{{Wang}}, \binits{J.}},
\bauthor{\bsnm{{Burford}}, \binits{J.}},
\bauthor{\bsnm{{Richardson}}, \binits{A.}},
\bauthor{\bsnm{{Rosebery}}, \binits{T.}},
\bauthor{\bsnm{{Rowe}}, \binits{J.}},
\bauthor{\bsnm{{Hogg}}, \binits{H.S.}}:
\bjtitle{\aj}
\bvolume{122}(\bissue{5}),
\bfpage{2587}
(\byear{2001})
\end{barticle}
\endbibitem

\bibitem[\protect\citeauthoryear{{Dworetsky}}{1983}]{Dworetsky1983}
\begin{barticle}
\bauthor{\bsnm{{Dworetsky}}, \binits{M.M.}}:
\bjtitle{\mnras}
\bvolume{203},
\bfpage{917}
(\byear{1983})
\end{barticle}
\endbibitem

\bibitem[\protect\citeauthoryear{{Fabricius} et~al.}{2021}]{Fabricius2021}
\begin{barticle}
\bauthor{\bsnm{{Fabricius}}, \binits{C.}},
\bauthor{\bsnm{{Luri}}, \binits{X.}},
\bauthor{\bsnm{{Arenou}}, \binits{F.}},
\bauthor{\bsnm{{Babusiaux}}, \binits{C.}},
\bauthor{\bsnm{{Helmi}}, \binits{A.}},
\bauthor{\bsnm{{Muraveva}}, \binits{T.}},
\bauthor{\bsnm{{Reyl{\'e}}}, \binits{C.}},
\bauthor{\bsnm{{Spoto}}, \binits{F.}},
\bauthor{\bsnm{{Vallenari}}, \binits{A.}},
\bauthor{\bsnm{{Antoja}}, \binits{T.}},
\bauthor{\bsnm{{Balbinot}}, \binits{E.}},
\bauthor{\bsnm{{Barache}}, \binits{C.}},
\bauthor{\bsnm{{Bauchet}}, \binits{N.}},
\bauthor{\bsnm{{Bragaglia}}, \binits{A.}},
\bauthor{\bsnm{{Busonero}}, \binits{D.}},
\bauthor{\bsnm{{Cantat-Gaudin}}, \binits{T.}},
\bauthor{\bsnm{{Carrasco}}, \binits{J.M.}},
\bauthor{\bsnm{{Diakit{\'e}}}, \binits{S.}},
\bauthor{\bsnm{{Fabrizio}}, \binits{M.}},
\bauthor{\bsnm{{Figueras}}, \binits{F.}},
\bauthor{\bsnm{{Garcia-Gutierrez}}, \binits{A.}},
\bauthor{\bsnm{{Garofalo}}, \binits{A.}},
\bauthor{\bsnm{{Jordi}}, \binits{C.}},
\bauthor{\bsnm{{Kervella}}, \binits{P.}},
\bauthor{\bsnm{{Khanna}}, \binits{S.}},
\bauthor{\bsnm{{Leclerc}}, \binits{N.}},
\bauthor{\bsnm{{Licata}}, \binits{E.}},
\bauthor{\bsnm{{Lambert}}, \binits{S.}},
\bauthor{\bsnm{{Marrese}}, \binits{P.M.}},
\bauthor{\bsnm{{Masip}}, \binits{A.}},
\bauthor{\bsnm{{Ramos}}, \binits{P.}},
\bauthor{\bsnm{{Robichon}}, \binits{N.}},
\bauthor{\bsnm{{Robin}}, \binits{A.C.}},
\bauthor{\bsnm{{Romero-G{\'o}mez}}, \binits{M.}},
\bauthor{\bsnm{{Rubele}}, \binits{S.}},
\bauthor{\bsnm{{Weiler}}, \binits{M.}}:
\bjtitle{\aap}
\bvolume{649},
\bfpage{5}
(\byear{2021})
\end{barticle}
\endbibitem

\bibitem[\protect\citeauthoryear{{Gaia Collaboration} et~al.}{2016}]{Gaia2016}
\begin{barticle}
\bauthor{\bsnm{{Gaia Collaboration}}},
\bauthor{\bsnm{{Prusti}}, \binits{T.}},
\bauthor{\bsnm{{de Bruijne}}, \binits{J.H.J.}},
\bauthor{\bparticle{et} \bsnm{al.}}:
\bjtitle{\aap}
\bvolume{595},
\bfpage{1}
(\byear{2016})
\end{barticle}
\endbibitem

\bibitem[\protect\citeauthoryear{{Gaia Collaboration} et~al.}{2023}]{Gaia2023}
\begin{barticle}
\bauthor{\bsnm{{Gaia Collaboration}}},
\bauthor{\bsnm{{Vallenari}}, \binits{A.}},
\bauthor{\bsnm{{Brown}}, \binits{A.G.A.}},
\bauthor{\bsnm{{Prusti}}, \binits{T.}},
\bauthor{\bparticle{et} \bsnm{al.}}:
\bjtitle{\aap}
\bvolume{674},
\bfpage{1}
(\byear{2023})
\end{barticle}
\endbibitem

\bibitem[\protect\citeauthoryear{{Harris}}{1996}]{Harris1996}
\begin{barticle}
\bauthor{\bsnm{{Harris}}, \binits{W.E.}}:
\bjtitle{\aj}
\bvolume{112},
\bfpage{1487}
(\byear{1996})
\end{barticle}
\endbibitem

\bibitem[\protect\citeauthoryear{{Kunder} et~al.}{2013a}]{Kunder2013a}
\begin{barticle}
\bauthor{\bsnm{{Kunder}}, \binits{A.}},
\bauthor{\bsnm{{Stetson}}, \binits{P.B.}},
\bauthor{\bsnm{{Cassisi}}, \binits{S.}},
\bauthor{\bsnm{{Layden}}, \binits{A.}},
\bauthor{\bsnm{{Bono}}, \binits{G.}},
\bauthor{\bsnm{{Catelan}}, \binits{M.}},
\bauthor{\bsnm{{Walker}}, \binits{A.R.}},
\bauthor{\bsnm{{Paredes Alvarez}}, \binits{L.}},
\bauthor{\bsnm{{Clem}}, \binits{J.L.}},
\bauthor{\bsnm{{Matsunaga}}, \binits{N.}},
\bauthor{\bsnm{{Salaris}}, \binits{M.}},
\bauthor{\bsnm{{Lee}}, \binits{J.-W.}},
\bauthor{\bsnm{{Chaboyer}}, \binits{B.}}:
\bjtitle{\aj}
\bvolume{146}(\bissue{5}),
\bfpage{119}
(\byear{2013}a)
\end{barticle}
\endbibitem

\bibitem[\protect\citeauthoryear{{Kunder} et~al.}{2013b}]{Kunder2013b}
\begin{barticle}
\bauthor{\bsnm{{Kunder}}, \binits{A.}},
\bauthor{\bsnm{{Stetson}}, \binits{P.B.}},
\bauthor{\bsnm{{Catelan}}, \binits{M.}},
\bauthor{\bsnm{{Walker}}, \binits{A.R.}},
\bauthor{\bsnm{{Amigo}}, \binits{P.}}:
\bjtitle{\aj}
\bvolume{145}(\bissue{2}),
\bfpage{33}
(\byear{2013}b)
\end{barticle}
\endbibitem

\bibitem[\protect\citeauthoryear{{Lenz} and {Breger}}{2005}]{Lenz2005}
\begin{barticle}
\bauthor{\bsnm{{Lenz}}, \binits{P.}},
\bauthor{\bsnm{{Breger}}, \binits{M.}}:
\bjtitle{Communications in Asteroseismology}
\bvolume{146},
\bfpage{53}
(\byear{2005})
\end{barticle}
\endbibitem

\bibitem[\protect\citeauthoryear{{Luna} et~al.}{2023}]{Luna2023}
\begin{botherref}
\oauthor{\bsnm{{Luna}}, \binits{A.}},
\oauthor{\bsnm{{Marchetti}}, \binits{T.}},
\oauthor{\bsnm{{Rejkuba}}, \binits{M.}},
\oauthor{\bsnm{{Minniti}}, \binits{D.}},
\oauthor{\bsnm{{.}}}:
arXiv e-prints,
2307
(2023).
\arxivurl{2307.13719}
\end{botherref}
\endbibitem

\bibitem[\protect\citeauthoryear{{McNamara}}{2000}]{McNamara2000}
\begin{barticle}
\bauthor{\bsnm{{McNamara}}, \binits{D.H.}}:
\bjtitle{\pasp}
\bvolume{112}(\bissue{774}),
\bfpage{1096}
(\byear{2000})
\end{barticle}
\endbibitem

\bibitem[\protect\citeauthoryear{{Netzel} et~al.}{2018}]{Netzel2018}
\begin{barticle}
\bauthor{\bsnm{{Netzel}}, \binits{H.}},
\bauthor{\bsnm{{Smolec}}, \binits{R.}},
\bauthor{\bsnm{{Soszy{\'n}ski}}, \binits{I.}},
\bauthor{\bsnm{{Udalski}}, \binits{A.}}:
\bjtitle{\mnras}
\bvolume{480}(\bissue{1}),
\bfpage{1229}
(\byear{2018})
\end{barticle}
\endbibitem

\bibitem[\protect\citeauthoryear{{Pols} et~al.}{1997}]{Pols1997}
\begin{barticle}
\bauthor{\bsnm{{Pols}}, \binits{O.R.}},
\bauthor{\bsnm{{Tout}}, \binits{C.A.}},
\bauthor{\bsnm{{Schroder}}, \binits{K.-P.}},
\bauthor{\bsnm{{Eggleton}}, \binits{P.P.}},
\bauthor{\bsnm{{Manners}}, \binits{J.}}:
\bjtitle{MNRAS}
\bvolume{289}(\bissue{4}),
\bfpage{869}
(\byear{1997})
\end{barticle}
\endbibitem

\bibitem[\protect\citeauthoryear{{Pols} et~al.}{1998}]{Pols1998}
\begin{barticle}
\bauthor{\bsnm{{Pols}}, \binits{O.R.}},
\bauthor{\bsnm{{Schr{\"o}der}}, \binits{K.-P.}},
\bauthor{\bsnm{{Hurley}}, \binits{J.R.}},
\bauthor{\bsnm{{Tout}}, \binits{C.A.}},
\bauthor{\bsnm{{Eggleton}}, \binits{P.P.}}:
\bjtitle{MNRAS}
\bvolume{298}(\bissue{2}),
\bfpage{525}
(\byear{1998})
\end{barticle}
\endbibitem

\bibitem[\protect\citeauthoryear{{Riello} et~al.}{2021}]{Riello2021}
\begin{barticle}
\bauthor{\bsnm{{Riello}}, \binits{M.}},
\bauthor{\bsnm{{De Angeli}}, \binits{F.}},
\bauthor{\bsnm{{Evans}}, \binits{D.W.}}, \betal:
\bjtitle{\aap}
\bvolume{649},
\bfpage{3}
(\byear{2021})
\end{barticle}
\endbibitem

\bibitem[\protect\citeauthoryear{{Rossi} et~al.}{2015}]{Rossi2015}
\begin{barticle}
\bauthor{\bsnm{{Rossi}}, \binits{L.J.}},
\bauthor{\bsnm{{Ortolani}}, \binits{S.}},
\bauthor{\bsnm{{Barbuy}}, \binits{B.}},
\bauthor{\bsnm{{Bica}}, \binits{E.}},
\bauthor{\bsnm{{Bonfanti}}, \binits{A.}}:
\bjtitle{\mnras}
\bvolume{450}(\bissue{3}),
\bfpage{3270}
(\byear{2015})
\end{barticle}
\endbibitem

\bibitem[\protect\citeauthoryear{{Schlafly} and
  {Finkbeiner}}{2011}]{Schlafly2011}
\begin{barticle}
\bauthor{\bsnm{{Schlafly}}, \binits{E.F.}},
\bauthor{\bsnm{{Finkbeiner}}, \binits{D.P.}}:
\bjtitle{\apj}
\bvolume{737},
\bfpage{103}
(\byear{2011})
\end{barticle}
\endbibitem

\bibitem[\protect\citeauthoryear{{Schlegel} et~al.}{1998}]{Schlegel1998}
\begin{barticle}
\bauthor{\bsnm{{Schlegel}}, \binits{D.J.}},
\bauthor{\bsnm{{Finkbeiner}}, \binits{D.P.}},
\bauthor{\bsnm{{Davis}}, \binits{M.}}:
\bjtitle{ApJ}
\bvolume{500},
\bfpage{525}
(\byear{1998})
\end{barticle}
\endbibitem

\bibitem[\protect\citeauthoryear{{Schr{\"o}der} et~al.}{1997}]{KPS1997}
\begin{barticle}
\bauthor{\bsnm{{Schr{\"o}der}}, \binits{K.-P.}},
\bauthor{\bsnm{{Pols}}, \binits{O.R.}},
\bauthor{\bsnm{{Eggleton}}, \binits{P.P.}}:
\bjtitle{MNRAS}
\bvolume{285}(\bissue{4}),
\bfpage{696}
(\byear{1997})
\end{barticle}
\endbibitem

\bibitem[\protect\citeauthoryear{{Soszy{\'n}ski} et~al.}{2013}]{Soszynski2013}
\begin{barticle}
\bauthor{\bsnm{{Soszy{\'n}ski}}, \binits{I.}},
\bauthor{\bsnm{{Udalski}}, \binits{A.}},
\bauthor{\bsnm{{Szyma{\'n}ski}}, \binits{M.K.}},
\bauthor{\bsnm{{Kubiak}}, \binits{M.}},
\bauthor{\bsnm{{Pietrzy{\'n}ski}}, \binits{G.}},
\bauthor{\bsnm{{Wyrzykowski}}, \binits{{\L}.}},
\bauthor{\bsnm{{Ulaczyk}}, \binits{K.}},
\bauthor{\bsnm{{Poleski}}, \binits{R.}},
\bauthor{\bsnm{{Koz{\l}owski}}, \binits{S.}},
\bauthor{\bsnm{{Pietrukowicz}}, \binits{P.}},
\bauthor{\bsnm{{Skowron}}, \binits{J.}}:
\bjtitle{\actaa}
\bvolume{63}(\bissue{1}),
\bfpage{21}
(\byear{2013})
\end{barticle}
\endbibitem

\bibitem[\protect\citeauthoryear{{Soszy{\'n}ski} et~al.}{2014}]{Soszynski2014}
\begin{barticle}
\bauthor{\bsnm{{Soszy{\'n}ski}}, \binits{I.}},
\bauthor{\bsnm{{Udalski}}, \binits{A.}},
\bauthor{\bsnm{{Szyma{\'n}ski}}, \binits{M.K.}},
\bauthor{\bsnm{{Pietrukowicz}}, \binits{P.}},
\bauthor{\bsnm{{Mr{\'o}z}}, \binits{P.}},
\bauthor{\bsnm{{Skowron}}, \binits{J.}},
\bauthor{\bsnm{{Koz{\l}owski}}, \binits{S.}},
\bauthor{\bsnm{{Poleski}}, \binits{R.}},
\bauthor{\bsnm{{Skowron}}, \binits{D.}},
\bauthor{\bsnm{{Pietrzy{\'n}ski}}, \binits{G.}},
\bauthor{\bsnm{{Wyrzykowski}}, \binits{L.}},
\bauthor{\bsnm{{Ulaczyk}}, \binits{K.}},
\bauthor{\bsnm{{Kubiak}}, \binits{M.}}:
\bjtitle{\actaa}
\bvolume{64}(\bissue{3}),
\bfpage{177}
(\byear{2014})
\end{barticle}
\endbibitem

\bibitem[\protect\citeauthoryear{{Spaenhauer} et~al.}{1992}]{Spaenhauer1992}
\begin{barticle}
\bauthor{\bsnm{{Spaenhauer}}, \binits{A.}},
\bauthor{\bsnm{{Jones}}, \binits{B.F.}},
\bauthor{\bsnm{{Whitford}}, \binits{A.E.}}:
\bjtitle{\aj}
\bvolume{103},
\bfpage{297}
(\byear{1992})
\end{barticle}
\endbibitem

\bibitem[\protect\citeauthoryear{{Terndrup} et~al.}{1998}]{Terndrup1998}
\begin{barticle}
\bauthor{\bsnm{{Terndrup}}, \binits{D.M.}},
\bauthor{\bsnm{{Popowski}}, \binits{P.}},
\bauthor{\bsnm{{Gould}}, \binits{A.}},
\bauthor{\bsnm{{Rich}}, \binits{R.M.}},
\bauthor{\bsnm{{Sadler}}, \binits{E.M.}}:
\bjtitle{\aj}
\bvolume{115}(\bissue{4}),
\bfpage{1476}
(\byear{1998})
\end{barticle}
\endbibitem

\bibitem[\protect\citeauthoryear{{Udalski} et~al.}{1992}]{Udalski1992}
\begin{barticle}
\bauthor{\bsnm{{Udalski}}, \binits{A.}},
\bauthor{\bsnm{{Szymanski}}, \binits{M.}},
\bauthor{\bsnm{{Kaluzny}}, \binits{J.}},
\bauthor{\bsnm{{Kubiak}}, \binits{M.}},
\bauthor{\bsnm{{Mateo}}, \binits{M.}}:
\bjtitle{\actaa}
\bvolume{42},
\bfpage{253}
(\byear{1992})
\end{barticle}
\endbibitem

\bibitem[\protect\citeauthoryear{{Yepez} et~al.}{2020}]{Yepez20}
\begin{barticle}
\bauthor{\bsnm{{Yepez}}, \binits{M.A.}},
\bauthor{\bsnm{{Arellano Ferro}}, \binits{A.}},
\bauthor{\bsnm{{Deras}}, \binits{D.}}:
\bjtitle{\mnras}
\bvolume{494}(\bissue{3}),
\bfpage{3212}
(\byear{2020})
\end{barticle}
\endbibitem

\bibitem[\protect\citeauthoryear{{Yepez} et~al.}{2022}]{Yepez2022}
\begin{barticle}
\bauthor{\bsnm{{Yepez}}, \binits{M.A.}},
\bauthor{\bsnm{{Arellano Ferro}}, \binits{A.}},
\bauthor{\bsnm{{Deras}}, \binits{D.}},
\bauthor{\bsnm{{Bustos Fierro}}, \binits{I.}},
\bauthor{\bsnm{{Muneer}}, \binits{S.}},
\bauthor{\bsnm{{Schr{\"o}der}}, \binits{K.-P.}}:
\bjtitle{\mnras}
\bvolume{511}(\bissue{1}),
\bfpage{1285}
(\byear{2022})
\end{barticle}
\endbibitem

\bibitem[\protect\citeauthoryear{{Zinn} and {West}}{1984}]{Zinn1984}
\begin{barticle}
\bauthor{\bsnm{{Zinn}}, \binits{R.}},
\bauthor{\bsnm{{West}}, \binits{M.J.}}:
\bjtitle{\apjs}
\bvolume{55},
\bfpage{45}
(\byear{1984})
\end{barticle}
\endbibitem

\end{thebibliography}

\appendix
\section{Comments on individual stars}
\label{sec:IND_STARS}

In this section we comment on the light curves, variable types and nature of some 
interesting or peculiar variables in Table \ref{variables} and Figs. \ref{RRLYR}, \ref{OSs} and \ref{nuevas}. We put some emphasis
on the amplitude and phase modulations of the Blazhko type  
in specific stars. In all the stars
labelled '$Bl$' in Table \ref{variables} the amplitude variations are neatly
distinguished in
the light curves of Figs. \ref{RRLYR}, particularly  in the  $I$-band.

{\bf V4, OG1}. The $I$-band light curves of these two RRab stars display very large difference between their OGLE III and OGLE IV data. In the paper by \citet{Soszynski2014} the authors attribute these larger differences to crowding and blending by unresolved stars.

{\bf V4, COG}. It has been a great confusion regarding the identification of the variable V4, discovered by \citep{Baade1946}. The description of a series of misunderstandings can be found in the notes on individual stas of NGC 6522 in the CVSGC \citep{cle01}. To make the story short,  consider that these two stars are separated by about 2 arcsec, hence their light curves are mutually contaminated. The star identified as V4 in the 2016 edition of the CVSGC, the latest at the time of writing this paper, is in fact the OGLE IV and $Gaia$ source
OGLE-BLG-RRLYR-12132 and 4050198106370354816 respectively. There is no available $Gaia$ light curve for this source, but OGLE IV data display a well developed $V$-band and $I$-band RRab-like light curve, with means at about 18.8 mag and 17.8 mag respectively, as shown with red dots in Fig. \ref{RRLYR}. This $Gaia$ source has no proper motions and hence its membership status is unclear. On the other hand, the $Gaia$ source 4050198110543863808  corresponds to a $V=16.5$ mag star that falls right on the HB and whose proper motions identify the star as a very likely cluster member. An independent period analysis of both the OGLE (OGLE-BLG-RRLYR-12132) and $Gaia$ (4050198110543863808) light curve render exactly the same period, confirming that we are in fact dealing with the same light source. 

We have been unable to find any old identification chart of V4, but in our opinion the authentic V4 is the one observed by $Gaia$ and consider that the OGLE IV identification is that of a nearby fainter  ($V \sim 18.8$ mag) stars whose apparent variability is the result of light pollution from the authentic V4, which in fact turns to appear as a truly cluster member RRab star. Note that OGLE III $I$-band light curve (black dots in Fig. \ref{RRLYR}) correctly corresponds to the authentic V4. The described identification problem seems to be exclusive of the OGLE IV light curve, which is likely due to crowding, as pointed out by \citet{Soszynski2014}.

In Table \ref{variables} we have listed the star V4 according to the above discussion. In the identification chart of Fig. \ref{CHART}, the true V4 is labeled whereas the OGLE-BLG-RRLYR-12132 stars is labeled as COG, the variability of which is but the reflection from V4.

{\bf V25}. Displays a well developed variation reasonably phased with a period of 1.462 days (Fig. \ref{nuevas}). This suggest the star to be a CW variable. However its $Gaia$ colour $BP-RP= 3.929$, or its transformation into $V-I=3.908$, puts the star off the CMD bounds in Fig. \ref{CMD_HB}. In spite of the differential reddening in NGC 6522, we believe this extreme reddening can only be of local rather circumstellar nature. We call attention to star V7, a non cluster member, equally largely reddened with $BP-RP= 4.359$.

{\bf V26}. The $G$ light curve of this star shows clear short term variations that can be phased with a period of  0.072922 days. The resulting light curve reminds that of an SX Phe star. However, in CMD the star appears among the bright reg giants. Given that this star is probably a cluster member, we believe  this is an semiregular or SR variable with rapid oscillations on top of a longer term and probably no periodic variation.  

{\bf Blazhko variables: V2, V10 and OG4.}
These three RRab stars display well developed phase and amplitude modulations of the Blazhko type. The more dense $I$-band data for a frequency analysis. In all cases, after removing the main fundamental period from the frequency spectrum, it leaves two clear symmetrical side lobes which are indicators of the Blazhko periodicity. The estimated values for V2, V10 and OG4
are $150.6 \pm 0.2$, $55.21 \pm 0.07$ and $274.2 \pm 0.4$ days respectively.

\end{document}